\documentclass{emulateapj}
\usepackage{color,amsmath}
\newcommand{\fAGN}{\ensuremath{f({\rm AGN})_{\rm MIR}}}

\begin{document}

\title{A controlled study of cold dust content in galaxies from {\it z}=0-2}
\author{Allison Kirkpatrick\altaffilmark{1,2}, Alexandra Pope\altaffilmark{1}, Anna Sajina\altaffilmark{3}, Daniel A. Dale\altaffilmark{4}, Tanio D\'{i}az-Santos\altaffilmark{5}, Christopher C. Hayward\altaffilmark{6,7}, Yong Shi\altaffilmark{8}, Rachel S. Somerville\altaffilmark{9}, Sabrina Stierwalt\altaffilmark{10}, Lee Armus\altaffilmark{11}, Jeyhan S. Kartaltepe\altaffilmark{12}, Dale D. Kocevski\altaffilmark{13}, Daniel H. McIntosh\altaffilmark{14}, David B. Sanders\altaffilmark{15}, Lin Yan\altaffilmark{16}}

\altaffiltext{1}{Department of Astronomy, University of Massachusetts, Amherst, MA 01002, USA}
\altaffiltext{2}{Yale Center for Astronomy \& Astrophysics, Physics Department, P.O. Box 208120, New Haven, CT 06520, USA, allison.kirkpatrick@yale.edu}
\altaffiltext{3}{Department of Physics \& Astronomy, Tufts University, Medford, MA 02155, USA}
\altaffiltext{4}{Department of Physics \& Astronomy, University of Wyoming, Laramie, WY 82071, USA}
\altaffiltext{5}{N\'{u}cleo de Astronom\'{i}a de la Facultad de Ingenier\'{i}a, Universidad Diego Portales, Av. Ej\'{e}rcito Libertador 441, Santiago, Chile}
\altaffiltext{6}{Center for Computational Astrophysics, Flatiron Institute, 162 Fifth Avenue, New York, NY 10010, USA}
\altaffiltext{7}{Harvard-Smithsonian Center for Astrophysics, 60 Garden Street, Cambridge, MA 02138, USA}
\altaffiltext{8}{School of Astronomy and Space Science, Nanjing University, Nanjing 210093, China}
\altaffiltext{9}{Department of Physics and Astronomy, Rutgers, The State University of New Jersey, 136 Frelinghuysen Rd, Piscataway, NJ 08854 USA}
\altaffiltext{10}{Department of Astronomy, University of Virginia, Charlottesville, VA 22904, USA}
\altaffiltext{11}{Spitzer Science Center, California Institute of Technology, Pasadena, CA 91125, USA}
\altaffiltext{12}{School of Physics and Astronomy, Rochester Institute of Technology, 84 Lomb Memorial Drive, Rochester, NY 14623, USA}
\altaffiltext{13}{Department of Physics and Astronomy, Colby College, Waterville, ME 04901, USA}
\altaffiltext{14}{Department of Physics and Astronomy, University of Missouri-Kansas City, 5110 Rockhill Road, Kansas City, MO 64110, USA}
\altaffiltext{15}{Institute for Astronomy, 2680 Woodlawn Drive, University of Hawaii, Honolulu, HI 96822, USA}
\altaffiltext{16}{Infrared Processing and Analysis Center, California Institute of Technology, Pasadena, CA 91125, USA}

\begin{abstract}
At $z=1-3$, the formation of new stars is dominated by dusty galaxies whose far-IR emission indicates they contain colder dust than local galaxies of a similar luminosity. We explore the reasons for the evolving IR emission of similar galaxies over cosmic time using: 1) Local galaxies from GOALS  ($L_{\rm IR}=10^{11}-10^{12}\,L_\odot$); 2) Galaxies at $z\sim0.1-0.5$ from the 5MUSES ($L_{\rm IR}=10^{10}-10^{12}\,L_\odot$); 3) IR luminous galaxies spanning $z=0.5-3$ from GOODS and {\it Spitzer} xFLS ($L_{\rm IR}>10^{11}\,L_\odot$). All samples have {\it Spitzer} mid-IR spectra, and {\it Herschel} and ground-based submillimeter imaging covering the full IR spectral energy distribution, allowing us to robustly measure $L_{\rm IR}^{\rm\scriptscriptstyle SF}$, $T_{\rm dust}$, and $M_{\rm dust}$ for every galaxy. Despite similar infrared luminosities, $z>0.5$ dusty star forming galaxies have a factor of 5 higher dust masses and 5\,K colder temperatures. The increase in dust mass is linked with an increase in the gas fractions with redshift, and we do not observe a similar increase in stellar mass or star formation efficiency. $L_{160}^{\rm\scriptscriptstyle SF}/L_{70}^{\rm\scriptscriptstyle SF}$, a proxy for $T_{\rm dust}$, is strongly correlated with $L_{\rm IR}^{\rm\scriptscriptstyle SF}/M_{\rm dust}$ independently of redshift. We measure merger classification and galaxy size for a subsample, and there is no obvious correlation between these parameters and $L_{\rm IR}^{\rm \scriptscriptstyle SF}/M_{\rm dust}$ or $L_{160}^{\rm\scriptscriptstyle SF}/L_{70}^{\rm\scriptscriptstyle SF}$. In dusty star forming galaxies, the change in $L_{\rm IR}^{\rm\scriptscriptstyle SF}/M_{\rm dust}$ can fully account for the observed colder dust temperatures, suggesting that any change in the spatial extent of the interstellar medium is a second order effect.\end{abstract}

\section{Introduction}
As the Universe ages, a phenomenon known as cosmic downsizing shifts the bulk of star formation to smaller galaxies. Additionally, as the Universe expands, the merger rate of galaxies decreases, and mergers become a less important triggering mechanism for star formation \citep{conselice2008,conselice2009,bluck2012}. These changes mean that, in certain respects, the composition of galaxies in the local Universe is quite different from what we would see if we existed in a galaxy at cosmic noon, 8-10 billion years ago ($z=1-2$), when the Universe was forming most of its stars \citep{madau2014}. Galaxies at cosmic noon appear to be more compact than their present day counterparts of similar mass \citep[e.g.,][]{trujillo2007,van2008,williams2014}, suggesting that the inner parts of galaxies are in place before the outer parts form \citep[e.g.,][]{hopkins2009,nelson2016}.

On the other hand, the cosmic noon Universe shares some key similarities with our present day Universe. For example, the backbone of the Hubble sequence was already in place by $z\sim2$, so that blue, disky galaxies were strongly star forming \citep[e.g.][]{lee2013}. Crucially, the main sequence of galaxy formation already existed by $z\sim3$ \citep[e.g.][]{behroozi2013,suzuki2015}. The main sequence is the tight empirical relationship between SFR and $M_\ast$ that defines the so-called ``normal mode" of secular star formation, as opposed to short-lived intense starbursts, typically triggered by mergers \citep[e.g.,][]{brinchmann2004,noeske2007,elbaz2011}. 

At cosmic noon, the galaxies dominating the buildup of stellar mass are massive, dusty, infrared luminous systems 
($L_{\rm IR}>10^{11}\,L_\odot$), but today, less massive, relatively isolated galaxies such as our Milky Way are the dominant sites of star formation \citep{murphy2011,madau2014}. IR luminous galaxies exist in the local Universe as well, but they contribute little to the current cosmic SFR density.
Historically, galaxy IR spectral energy distributions (SEDs) were parameterized only by $L_{\rm IR}$ \citep{sanders1996,dale2001,chary2001,rieke2009}. 
However, {\it Herschel} brought new insight into the peak of the SED in distant galaxies, and we now know that at $z\sim1-2$, Ultra Luminous Infrared Galaxies (ULIRGs) have colder far-IR emission than galaxies with the same luminosity at $z\sim0$ \citep{chapman2002,pope2006,sajina2006,clements2010,nordon2010,symeonidis2009,magnelli2011,elbaz2011,sajina2012,kirkpatrick2012}.  

The shape of the ULIRG SED evolves strongly from $z\sim0-2$, but local ULIRGs represent the most extreme objects. They are merging, compact starbursts, which are heavily obscured and even optically 
thick at far-IR wavelengths \citep[e.g.][]{sakamoto2008,scoville2015b}. 
These objects also lie well above the main sequence, while ULIRGs at $z\sim1-2$ lie on the main sequence, indicative of a slower, secular star formation history \citep{elbaz2011}. However, local LIRGs, which are an order of magnitude less luminous than local ULIRGs, share certain dust emission characteristics with their high redshift luminosity-based counterparts indicating they may have similar ISMs despite the fact that local LIRGs also lie above the main sequence. For example, $z\sim1-2$ ULIRGs and LIRGs have similar polycyclic aromatic hydrocarbon (PAH) emission strength, parameterized by $6.2\,\mu$m equivalent widths and $L_{6.2}/L_{\rm IR}$, as local LIRGs \citep{pope2013,stierwalt2014,kirkpatrick2014}. The similarity of these features indicates a similar photodissociation region structure in these galaxies, hinting at similarities between their interstellar media \citep{pope2013,stierwalt2014}. Therefore, despite evolution of the main sequence with lookback time, it is reasonable to examine how closely dusty galaxies resemble each other from $z=0-2$.

Far-IR/submm emission is sensitive to dust temperature, emissivity, dust mass, the geometry of the star forming regions within the ISM, and the incident radiation field \citep{witt2000,gordon2001,misselt2001}, although the far-IR SED is most commonly used to calculate three parameters: $L_{\rm IR} (8-1000\,\mu$m); $M_{\rm dust}$, measured from the Rayleigh-Jeans tail; and $T_{\rm dust}$, which characterizes the peak of the SED. The change of SED shape, along with the similarity and differences between the distant and local Universe, leads to a key question in galaxy evolution: 
What is the root cause of the colder dust emission at $z\sim1-2$?

Recently, it has been suggested that $L_{\rm IR}$ {\it and} ISM geometry (including surface density and compactness) fully account for the shape of the SED; that is, high redshift IR luminous galaxies are colder because they are more extended \citep{elbaz2011,rujopakarn2013}. Additionally, the overall structure of the ISM in high redshift galaxies may be different.  A larger fraction of gas is likely in H$_2$ rather than H{\sc i}, the mass function of giant molecular clouds evolves, and proportionally more of the dust may arise from the diffuse ISM in distant galaxies \citep{popping2014a,popping2014b,scoville2015}. However, theoretical simulations and radiative transfer calculations contend that $L_{\rm IR}$ and $M_{\rm dust}$ or total mass, $M$, are the primary parameters for determining a galaxy's global SED \citep{chakrabarti2005,saf2015}. Then, compactness of the ISM, mass function of molecular clouds, and possibly surface density, have second order effects over the shape of the SED. Indeed, \citet{lee2016} demonstrate at $z=0$ that heavily dust obscured galaxies with $L_{\rm IR} = 10^{11}-10^{12}\,L_\odot$ have similar dust masses and temperatures as their less obscured counterparts, even though the extremely dust obscured galaxies are likely much more compact. This would suggest that the geometry of the ISM in IR luminous galaxies has little overall effect on the measured dust mass and temperature. % is not evolving with redshift, but that IR luminous galaxies at $z\sim1-2$ are simply scaled up counterparts of less luminous local galaxies.

In this work, we test how the dust temperatures and dust masses change with $L_{\rm IR}$ and redshift for IR luminous galaxies from $z\sim0-2$. We utilize three extensively studied samples from the literature, which are large ($\gtrsim200$ galaxies in each sample) and have deep multiwavelength data sets that eliminate any need for photometric stacking. We analyze all the SEDs in a self-consistent manner. Crucially, all of our galaxies have mid-IR spectroscopy, which we use to remove AGN and compact starbursts, which excludes a potential source of uncertainty when comparing dust masses and temperatures \citep[e.g.][]{lee2016}.
In Section 2, we describe our three data sets and summarize the relevant literature pertaining to these samples; in Section 3, we discuss how we remove AGN and measure dust masses self-consistently; in Section 4, we characterize the shape of the far-IR SED using broadband colors; in Section 5, we measure an increase in the dust mass of DSFGs with redshift and parameterize the relationship between $L_{\rm IR}/M_{\rm dust}$ and $T_{\rm dust}$; in Section 6, we discuss the origin of the increased dust masses with redshift and whether ISM size or merger classification can influence the SED, and we summarize in Section 7. Throughout this work, we assume a flat cosmology with $H_{0}=70\,\rm{km}\,\rm{s}^{-1}\,\rm{Mpc}^{-1}$, $\Omega_{\rm{M}}=0.3$, and $\Omega_{\Lambda}=0.7$.

\section{Sample Description}
\label{sec:sample}
\begin{figure*}[ht!]
\centering
\includegraphics[width=6.5in]{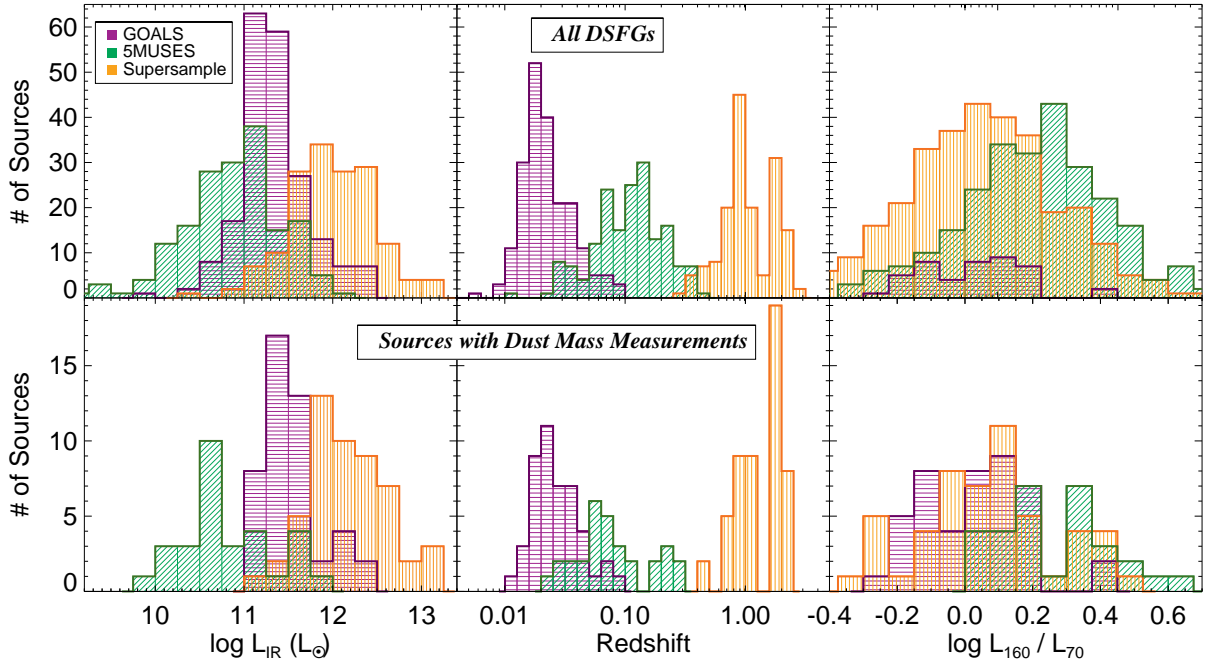}
\caption{$L_{\rm IR}$ ({\it left}), redshift ({\it middle}), and $L_{160}/L_{70} (\propto T_{\rm dust})$ ({\it right}) distributions of the dusty star forming galaxies (DSFGs) in the GOALS, 5MUSES, and Supersample. We include all sources with {\it Spitzer} IRS spectroscopy and $\fAGN<0.5$ (see Section \ref{sec:agn}) in the upper panels. The lower panels show the distributions for sources in which we were able to measure the dust mass (see Section \ref{sec:dust}).\label{z_dist}}
\end{figure*}

In this study, we compare the far-IR/submm properties of three samples at different redshifts extensively described in the literature (Figure \ref{z_dist}): 1) $z<0.088$: Great Observatories All Sky Survey \citep[GOALS; PI L.\,Armus ][]{armus2009}; 2) $z=0.05-0.75$: 5 mJy Unbiased {\it Spitzer} Extragalactic Survey
\citep[5MUSES; P.I. G.\,Helou][]{wu2010}; 3) $z=0.3-3.0$: {\it Spitzer} IRS Supersample \citep{kirkpatrick2015}. These samples comprise a case study of dusty star forming galaxies where the far-IR SEDs are well sampled and stacking is not required. Indeed, these samples are ideal for comparison as they each contain massive ($M_\ast \gtrsim 10^{9}\,M_\odot$), infrared luminous ($L_{\rm IR} = 10^{10}-10^{13}\,L_\odot$) galaxies with far-IR imaging from the {\it Spitzer Space Telescope} and the {\it Herschel Space Observatory} and mid-IR spectroscopy from the {\it Spitzer} IRS instrument \citep{houck2004}, useful for quantifying and removing dust heating due to an obscured AGN. We briefly summarize the basic properties of each sample in the following subsections, and in Section 3, we discuss how we arrive at the final sample sizes of 46 galaxies from GOALS, 30 galaxies from 5MUSES, and 51 galaxies from the Supersample.

\subsection{GOALS}
The GOALS sample is comprised of 180 luminous IR galaxies (LIRGs; $L_{\rm IR}=10^{11}\,L_\odot$) and 22 ultra luminous IR galaxies (ULIRGs; $L_{\rm IR}>10^{12}L_\odot$). Many of these systems are interacting, resulting in 244 individual galactic nuclei with {\it Spitzer} IRS spectroscopy. These galaxies are a representative subset of the IRAS Bright Galaxy Survey \citep{sanders2003} and were selected to have $S_{60} > 5.24\,$Jy. GOALS sources cover the distance range 15 Mpc $<$ D $<$ 400 Mpc, which corresponds to $z<0.088$.

The IRS spectra have been previously analyzed in depth \citep{diaz2010,diaz2011,u2012,stierwalt2014}. Here, we make use  of the IRS SL staring observations, $\lambda=5.5-14.5\,\mu$m, to identify and remove AGN from the sample (Section \ref{sec:agn}).
The GOALS galaxies have global flux densities from all three {\it Spitzer} MIPS bandpasses (Mazzarella et al.\,2017, in prep). $L_{\rm IR} (8-1000\,\mu$m) was calculated following the formula in \citet{sanders1996} using all four {\it IRAS} flux densities \citep{diaz2010}. 
%Analysis of {\it HST}/ACS imaging for a fraction of the sample is presented in \citet{haan2010}, \citet{kim2013}. 

The UV-IR SEDs were analyzed in detail for a subset of 64 galaxies \citep[see][for a discussion of the selection of these objects]{u2012}. These galaxies have stellar mass measurements, so they form our initial dust mass sample as well. From the 64 galaxies, 58 sources have 850\,$\mu$m from the {\it James Clerk Maxwell Telescope} or 250\,$\mu$m photometry from {\it Herschel} \citep{chu2017}, which we use to calculate $M_{\rm dust}$. Five of these are double nuclei sources that we remove, and seven are hosting AGN. This leaves a final sample size of 46 galaxies.

\subsection{5MUSES}
5MUSES is a {\it Spitzer} IRS mid-IR spectroscopic survey of 330 galaxies selected from the SWIRE and {\it Spitzer} Extragalactic First Look Survey fields \citep[details in][]{wu2010}. It is a flux limited sample selected at 24\,$\mu$m, with $S_{24}>5\,$mJy. 280 sources have optical spectroscopic redshifts \citep{wu2010}. 5MUSES is a representative sample at intermediate redshift (the median redshift of the sample is 0.14) of galaxies with $L_{\rm IR}\sim10^{10}-10^{12}\,L_\odot$, bridging the gap between local (U)LIRGs and high redshift observations.

For the present study, we make use of the {\it Spitzer} IRS SL spectroscopy to quantify AGN emission. Complete details of the {\it Spitzer} data reduction are found in \citet{wu2010}. IRS spectra combined with {\it Spitzer} MIPS imaging is used to calculate $L_{\rm IR}(5-1000\,\mu$m) in \citet{wu2010}, with a slightly different cosmology ($\Omega_M=0.27, \Lambda=0.73$). We adjust the published $L_{\rm IR}$s by multiplying by 0.99 to account for the difference in cosmology and 0.95 to scale to $L_{\rm IR}(8-1000\,\mu$m), which we have determined using a purely star forming template from \citet{kirkpatrick2015}. %The 0.95 multiplication factor underestimates what is needed if the galaxy hosts a luminous AGN, but as we focus primarily on the emission of SFGs, we do not further adjust the $L_{\rm IR}$ of AGN.
We use {\it Herschel} SPIRE observations of a subset of 188 sources in the {\it Herschel} Multi-Tiered Extragalactic Suvery \citep[HerMES][]{oliver2010,oliver2012,magdis2013} to calculate $M_{\rm dust}$.

Of the 188 {\it Herschel} galaxies, 40 have 500\,$\mu$m detections, necessary to calculate $M_{\rm dust}$. Six of these are removed because they host AGN. We remove an additional 4 sources where the 500\,$\mu$m photometry appears blended with a nearby source. This leaves a final sample of 30 galaxies.
%\citet{u2012} fit the GOALS galaxies with both a Salpeter and Chabrier IMF, and find an offset of 
%$M_\ast^{\rm Sal} = 0.55\,M_\ast^{\rm Cha}$, which we use to convert the 5MUSES $M_\ast$ to a Salpeter IMF.

\subsection{Supersample}
We have assembled a multi-wavelength data set for a sample of 343 high redshift ($z\sim0.3-4.0$) (U)LIRGs in the Great Observatories Origins Deep Survey North 
(GOODS-N),
Extended {\it Chandra} Deep Field Survey (ECDFS), and {\it Spitzer} xFLS fields. All sources are selected to have mid-IR spectroscopy from {\it Spitzer} IRS, which is used to measure the redshift \citep{kirkpatrick2012}. Our sample contains a range of sources from individual observing programs, each with differing selection criteria. However, the overarching selection criterion is a 24\,$\mu$m flux limit of 0.9\,mJy for the xFLS galaxies, which are taken from a shallower survey, and 0.1\,mJy for the GOODS and ECDFS galaxies, compiled from a deep survey. In addition to IRS spectra, these sources all have {\it Spitzer} MIPS and {\it Herschel} PACS and SPIRE imaging \citep{sajina2012,kirkpatrick2015}.
%The xFLS sample is described in detail in \citet{sajina2012}. 
%The sources were selected to have $S_{24}>0.9$\,mJy and to have an R magnitude of $m_{R,{\rm Vega}}\geq 20$. The xFLS IRS sample contains just under half of the xFLS sources that meet the above photometric criteria; however, \citet{sajina2012} find that the IRS sample is representative of a 24\,$\mu$m-selected sample ($>$\,0.9\,mJy) at $z$\,$\gtrsim$1.
%The GOODS-N and ECDFS samples include all sources in these fields that were observed with {\it Spitzer} IRS \citep[complete details in][]{kirkpatrick2012}. All of these sources were selected at 24\,$\mu$m, and 93\% of have $S_{24}>100\,\mu$Jy.
We calculate individual $L_{\rm IR}(8-1000\,\mu$m) by fitting a full suite of {\it Spitzer} and {\it Herschel} photometry with the library of templates from \citet{kirkpatrick2015}. Redshifts are determined by fitting the main PAH emission features, or using optical spectroscopy in the case of a featureless MIR spectrum. 

In this work, we also utilize 870\,$\mu$m photometry of GOODS-S from LABOCA on APEX \citep{weiss2009}, the combined AzTEC+MAMBO 1.1mm map of GOODS-N \citep{penner2011}, and MAMBO 1.2\,mm imaging of xFLS \citep{lutz2005,sajina2008,martinez2009}, which exists for 169 galaxies. For 4 of these, the source redshift is high enough that there is no data beyond $\lambda=250\,\mu$m, which is our threshold for calculating dust mass. We also remove any sources from the sample where the submm emission is blended with a nearby source (10 sources). We remove 57 galaxies for hosting an AGN. Finally, we remove 47 galaxies for have submm photometry with a SNR $<$ 2, leaving a final sample of 51 galaxies.

\subsection{Selection Effects}
We are comparing the far-IR/submm properties of three different samples with different selection criteria, so we must understand whether our results are biased due to the selection criteria. 
The GOALS sample is selected at 60\,$\mu$m while 5MUSES and the Supersample are selected at observed frame 24\,$\mu$m; this selection criteria alone could produce samples of galaxies that are warmer or colder for the same $L_{\rm IR}$ range.

We consider the 5MUSES and GOALS sources together to make up our low $z$ sample. All of the 5MUSES sources with $L_{\rm IR}>10^{11}$ (same as GOALS) would be selected at rest frame $S_{60}>5.24\,$Jy, the GOALS selection criterion. Similarly, all GOALS galaxies meet the 5MUSES selection criterion of $S_{24}>5\,$mJy.

We use the mid-IR spectrum to determine if a 5MUSES source meets the Supersample selection criterion, since at $z=1-2$, $S_{24}$ covers the PAH features. Only 28\% of the 5MUSES sample would be selected at $z=1$, and 11\% at $z=2$. This is not surprising, since the 5MUSES galaxies are on average less luminous. Only 32\% of the 5MUSES galaxies overlap in the $L_{\rm IR}$ range spanned by the Supersample. However, of the 60 5MUSES galaxies with $L_{\rm IR}>2\times 10^{11}\,L_\odot$ (the same range as the bulk of the Supersample), 50 would be selected as Supersample galaxies at $z=1$ and 30 would be selected at $z=2$, due to similar $L_{\rm PAH}/L_{\rm IR}$ ratios as $z\sim1-2$ (U)LIRGs \citep{pope2013,kirkpatrick2014,battisti2015}. Estimating whether the GOALS sources would be included in our Supersample is more nuanced, since we lack global IRS spectra for the GOALS galaxies. Instead, we use the \citet{chary2001} library of templates which were derived from the IRAS BGS. At $z=1$, templates with $\log L_{\rm IR}>11.11L_\odot$ meet the selection threshold, but at $z=2$, this increases to $\log L_{\rm IR}>12.21 L_\odot$. 

 %, which could result in different $T_{\rm dust}$ for samples with similar $L_{\rm IR}$ ranges, where $T_{\rm dust}$ characterizes the peak of the SED. 
Next, we estimate what fraction of the Supersample would be selected as a 5MUSES or GOALS galaxy, if the Supersample existed at a different redshift. This is largely an academic exercise, since the Supersample 
contains the most luminous galaxies, and \citet{symeonidis2010} demonstrated that the IRAS selection criteria are sensitive to ULIRGs ($L_{\rm IR}>10^{12}\,L_\odot$) with $T=17-87\,$K. Nevertheless, such a check will ensure that the Supersample contains enough warm dust to be luminous at rest frame 60$\mu$m and 24\,$\mu$m.
%First, we isolate Supersample sources in redshift ranges where either 100, 160, or 250\,$\mu$m falls in the rest frame range $\lambda=50-75\,\mu$m, approximately the bandpass of the IRAS 60\,$\mu$m filter. As this is a rough estimate, we do not apply any scaling to correct for the different wavelength ranges covered by the 100, 160, or 250\,$\mu$m bandpasses. We then calculate the threshold $L_{60}$ at $z=0.015$ for $S_{60}=5.24\,$Jy, the selection threshold of the GOALS survey. If $L_\nu+dL_\nu$ (the observed flux and uncertainty) is greater than this threshold, we count our Supersample source as a detection. 100\% of the 142 sources in the appropriate redshift range are detected. Second, 
We calculate synthetic 60\,$\mu$m photometry following the prescription in Section \ref{sec:syn}, for a redshift of $z=0.015$, which is the peak of the GOALS redshift distribution in Figure \ref{z_dist}. 97\% of the Supersample meets the GOALS selection criterion.
The 5MUSES selection criterion is $S_{24}>5\,$mJy. We shift our synthetic rest frame photometry to $z=0.15$ by scaling down by 1.62, a ratio determined with the \citet{kirkpatrick2015} template library. Again, 97\% would be selected as a 5MUSES galaxy. 
%We can check if 5MUSES sources meet the GOALS selection criteria using MIPS 70\,$\mu$m, since at $z=0.167$ (approximately the mean redshift of the 5MUSES galaxies), this corresponds to 60\,$\mu$m. Only 59\% meet the selection threshold. However, most of these are lower luminosity sources. When we restrict ourselves to sources with $L_{\rm IR}>10^{11}$, the percentage of selected sources increases to 99\%. 
%We can estimate directly whether the 5MUSES galaxies would be part of the Supersample utilizing their IRS spectra (24\,$\mu$m selection criteria covers PAH features at $z\sim1-2$). We redshift the spectra to $z=1$ and $z=2$ and calculate the observed frame 24\,$\mu$m flux density, where $S_{24}=100\,\mu$Jy is the selection threshold. Only 28\% of the 5MUSES sample would be selected at $z=1$, and 11\% at $z=2$. This is not surprising, since the 5MUSES galaxies are on average less luminous. Only 32\% of the 5MUSES galaxies overlap in the $L_{\rm IR}$ range spanned by the Supersample. Of the overlapping galaxies, most are selected as Supersample galaxies at $z=1$ due to similar $L_{\rm PAH}/L_{\rm IR}$ ratios as $z\sim1-2$ (U)LIRGs \citep{pope2013,kirkpatrick2014,battisti2015}. %If we restrict our selection to sources with $L_{\rm IR}> 10^{11}\,L_\odot$, then 56\% are selected at $z=1$. 

From this, we conclude that if cold, massive galaxies, like most of the Supersample, exist at lower redshift, they are easily detectable in current surveys.
Similarly, at $z\sim1$ at least, we should be able to select a population of galaxies with the range of dust masses and temperatures exhibited by local LIRGs, if they exist. However, at $z\sim2$, galaxies with spectral shapes similar to local LIRGs are undetectable in current 24\,$\mu$m surveys.
%The reader should bear in mind that the low $z$ samples are not biased against galaxies with $T_{\rm dust}$ and $M_{\rm dust}$ as the high redhisft sample, but the same does not hold true in reverse.

\subsection{ULIRG nomenclature}
Throughout the remainder of this paper, we dispense with the (U)LIRG nomenclature for clarity. In the local Universe, the term ULIRG evokes a massive merging system, where the merger gives rise to a starburst followed by an AGN \citep{sanders1996}. At earlier times, the correlation between ULIRG and merger still exists, but is more tenuous, and the link between merger and starburst is even less clear in these systems \citep{hopkins2010,kartaltepe2012,hayward2013}. 
LIRG and ULIRG are simply luminosity cuts, so to avoid any preconceived notions between these luminosity cuts and physical properties, we follow the convention in \citet{casey2014} and refer to objects in all three samples as dusty star forming galaxies (DSFG).

\section{Data and Physical Properties}
In this paper, we compare three samples of IR luminous galaxies at three different epochs. For a fair comparison, we shift all IR photometry to the rest frame, we remove any AGN contribution using {\it Spitzer} spectroscopy, and we derive dust masses self consistently. By using the mid-IR spectrum to look for a hot dust continuum, which we attribute to an AGN, we are also removing galaxies with a hot central starburst.%The galaxies that we analyze in the remainder of this paper all have similar 6.2 and 7.7\,$\mu$m strengths and $L_{\rm PAH}/L_{\rm IR}$ ratios\citep{stierwalt2013,pope2013,kirkpatrick2014}.

\subsection{Identifying and Removing AGN}
\label{sec:agn}
The mid-IR spectrum ($\lambda=3-20\,\mu$m) is rich for identifying star formation features and AGN. The most prominent dust emission complexes are produced by polycyclic aromatic hydrocarbons 
(PAHs, $\lambda=6.2,7.7,11.3,12.7\,\mu$m) which are abundant in galaxies with metallicity close to solar, such as high redshift DSFGs \citep{magdis2012}. These molecules are preferentially located in the photodissociation regions surrounding star forming regions \citep{helou2004}. As such, PAHs are good tracers of ongoing star formation in a galaxy \citep{peeters2004}. However, the mid-IR spectrum can also exhibit continuum emission due to the torus enveloping the AGN.

We perform spectral decomposition of the mid-IR spectrum ($\sim5-15\,\mu$m rest frame) for each source in all three samples in order to disentangle the AGN and star forming components. \citet{pope2008a} explain the technique in detail, and we summarize it here. We fit the individual spectra with a model comprised of four
components: (1) the star formation component is represented by the mid-IR spectrum of the prototypical starburst M~82; (2) the AGN component is determined by fitting a pure power-law ($\lambda^\alpha$) with the slope and 
normalization ($N$) as free parameters; (3,4) extinction curves from the \citet{weingartner2001} dust models for Milky Way (MW) type dust is applied to the AGN component and star forming component. The full model is then
\begin{equation}
\label{eq:decomp}
S_\nu=N_{\rm AGN}\lambda^\alpha\,e^{-\tau_{\rm AGN}}+N_{\rm \scriptscriptstyle SF} S_\nu ({\rm M82})e^{-\tau_{\rm \scriptscriptstyle SF}}
\end{equation}
We fit for the $N_{\rm AGN}$, $N_{\rm \scriptscriptstyle SF}$, $\alpha$, $\tau_{\rm AGN}$, and $\tau_{\rm \scriptscriptstyle SF}$ simultaneously.

For each source, we quantify the strength of the AGN, \fAGN, as the fraction of the total mid-IR
luminosity ($\lambda=5-15\,\mu$m) coming from the power-law continuum component. For the GOALS sample, our AGN identification technique selects the same sources as the $S_{15}/S_{5.5}$ v. $S_{6.2}/S_{5.5}$ diagnostic applied in \citet{petric2011}. For the 5MUSES sources, we also identify the same AGN as \citet{wu2010}, where the authors classify AGN on the basis of $6.2\,\mu$m PAH feature equivalent width alone. 

In \citet{kirkpatrick2015}, we found that an AGN had a significant contribution to the far-IR emission, quantified through dust temperatures and luminosities, when $\fAGN\geq0.5$. Recently, \citet{lee2016} showed that local dust obscured galaxies hosting an AGN had lower dust masses than their star forming counterparts. Consequently, we remove all 5MUSES, GOALS, and Supersample galaxies with $\fAGN\geq0.5$, so that we are only comparing the far-IR/submillimeter properties of strongly star forming systems. We note that because the GOALS sample is nearby, in some cases IRS observations only cover the central region of the galaxy, in contrast to the IRS observations of 5MUSES and the high $z$ Supersample. This could introduce a slight bias as we may remove sources that have a nucleus dominated by AGN emission, whereas the galaxy integrated emission is dominated by star formation. We only remove 13\% of the GOALS sample, so the effects of this potential bias are small.

Our AGN identification technique only relies on detecting hot continuum emission in the mid-IR. While we attribute this to an AGN, in principle, a very compact, dust enshrouded starburst would display the same mid-IR signature. By removing AGN, we are also then removing these compact starbursts, which will have different ISM properties than galaxies where the star forming regions are located throughout the galaxy, due to higher gas densities and a harsher radiation field. However, we note that although such compact starbursts exist locally \citep[e.g.][]{diaz2011}, so far such galaxies are rare at high redshift and still have prominent PAH features \citep{nelson2014}. From the Supersample, we remove 47\% for having $\fAGN\geq0.5$, and 62\% of these removed sources have X-ray detections consistent with being an AGN \citep{kirkpatrick2015}, while the remaining sources may be Compton thick based on predicted X-ray luminosities \citep{kirkpatrick2012}. As such, we do not think that we are missing a large population of compact starbursts by removing sources with $\fAGN\geq0.5$.

\subsection{Calculating Rest Frame Photometry}
\label{sec:syn}
In Section \ref{sec:shape}, we compare the far-IR SED shape of the GOALS, 5MUSES, and Supersample utilizing far-IR colors. To make a fair comparison between three samples at different redshift ranges, we require synthetic rest frame photometry in the MIPS and SPIRE photometric bandpasses for the 5MUSES and Supersample. GOALS is at low enough redshift that the offset in rest frame wavelengths is negligible. We fit templates to the Supersample and 5MUSES from the MIR-based Library in \citet{kirkpatrick2015}. The MIR-based Library is a suite of 11 empirical templates that characterize the full shape of the IR SED based on the relative amounts of PAH and continuum emission in a source's mid-IR spectroscopy. We select the appropriate template for each source from the spectral decomposition described above, fit to the far-IR photometry ($\lambda>20\,\mu$m), and then convolve each template with the MIPS and SPIRE transmission filters to create synthetic rest frame photometry at 70, 160, and 250\,$\mu$m. We plot the Supersample intrinsic SEDs and estimated rest frame photometry in Appendix \ref{appA}. The biggest concern when interpolating between data points is whether we have chosen a template with the correct far-IR shape. However, the majority of galaxies have a sufficiently well-sampled far-IR SED to mitigate this uncertainty, as the estimated photometry lies very close to observed data points.

It is unlikely that an AGN that is not significantly affecting the mid-IR emission would have a strong effect on the far-IR emission. Even so, we correct the rest-frame broadband colors and $L_{\rm IR}$s of all SFGs with $0.0<\fAGN<0.5$ in order to account for any scatter introduced by a buried AGN that may contribute to, but not dominate, the IR luminosity. We have calculated these corrections by decomposing empirical templates into their relative star formation and AGN components as described in detail in \citet{kirkpatrick2015}. The AGN corrections are listed below, for the rest frame luminosities. In \citet{kirkpatrick2015}, we discuss ways to estimate \fAGN\ in the absence of mid-IR spectroscopy, so these corrections can be applied to sources with only broadband photometry.
\begin{align}
\label{eq:corr}
L_{\rm IR}^{\rm \scriptscriptstyle SF} &= L_{\rm IR}\times(1+0.035\fAGN-0.66\fAGN^2) \nonumber\\
L_{24}^{\rm \scriptscriptstyle SF} &= L_{24}\times(1-1.81\fAGN+0.94\fAGN^2) \nonumber\\
L_{70}^{\rm \scriptscriptstyle SF}&=L_{70}\times(1-0.08\fAGN-0.16\fAGN^2)\nonumber\\
L_{160}^{\rm \scriptscriptstyle SF}&=L_{160}\times(1-0.036\fAGN)\nonumber\\
L_{250}^{\rm \scriptscriptstyle SF}&=L_{250}\times(1-0.018\fAGN)
\end{align}

\subsection{Dust Mass}
\label{sec:dust}
With our three samples, we can compare how $M_{\rm dust}$ evolves in dusty galaxy populations with redshift. Dust mass is calculated as
\begin{equation}
\label{eq:dust}
M_{\rm dust} = \frac{S_\nu D_L^2}{\kappa_\nu B_\nu(T)}
\end{equation}
where $B_\nu(T)$ is the Planck equation, $\kappa_\nu$ is the dust opacity, $D_L$ is the luminosity distance, and $S_\nu$ is the flux density. For $S_\nu$, we use the longest far-IR/submm data available, which varies with each sample. 
We only use sources whose rest frame measurement is $\lambda\geq250\,\mu$m, otherwise the dust emission may not be tracing the coldest dust component \citep{scoville2015}. When we combine this criterion with the $\fAGN<0.5$ criterion, we have the final sample sizes of 46 galaxies from GOALS, 30 galaxies from 5MUSES, and 52 galaxies from the Supersample.

We take $\kappa_\nu$, the dust opacity, from the \citet{weingartner2001}\footnote{Recent results indicate a change in these model opacities \citep{dalcanton2015,planck2016}, but we are consistently using the same model for every source, so we do not expect any change in opacity to affect our dust mass comparison}, models at the rest wavelength of $S_\nu$ for each individual source. We assume MW-like dust and $R_V=3.1$, although in these models, the far-IR/submm opacities have negligible changes for different $R_V$ and are similar for the MW-, LMC-, and SMC-type models (at 850\,$\mu$m, the opacity differs by $<$10\%). 

\citet{scoville2015} argue for using fixed $T_{\rm cold}$, since presumably the temperature in the diffuse ISM should be relatively similar for all massive, dusty galaxies, assuming similar radiation fields and dust grain mixtures. Indeed, in \citet{kirkpatrick2015} we found that the average cold dust temperature in SFGs in the Supersample was remarkably constant, $T\sim26\,$K, and did not scale with $L_{\rm IR}$. 
Accordingly, we set $T_{\rm cold}=25\,$K for all three samples and derive $M_{\rm dust}$ using Equation \ref{eq:dust}. As a test of this assumption, we fit each galaxy with a two temperature modified blackbody with $T_{\rm cold}=25\,$K and achieve excellent fits for all sources.
We explore how the dust masses change if $T_{\rm dust}$ is allowed to vary in Appendix \ref{appB} and find that all conclusions in this paper hold regardless of the specific method of measuring dust mass as long as these measurements are consistent across all three samples. In this study, we are primarily concerned with comparing the dust masses at high redshift relative to low redshift. By fixing $T_{\rm cold}$ and using the same $\kappa$ model for all samples, we are able to probe how dust masses change with redshift if the other dust properties of these galaxies remain the same.

\begin{deluxetable*}{lccc lcc}
\tablecolumns{7}
\tablecaption{Supersample Properties\label{properties}}
\tablehead{\colhead{Name} & \colhead{RA} & \colhead{Dec} & \colhead{\it z} & \colhead{$\log L_{\rm IR} (L_\odot)$\tablenotemark{a}} & \colhead{$\log M_{\rm dust} (M_\odot)$} & \colhead{$\log M_\ast (M_\odot)$\tablenotemark{b}}}%& %\colhead{$\lambda_{\rm dust}$\tablenotemark{b}}}
\startdata
GN\_IRS2 & 12:37:02.74 & +62:14:01.0 & 1.24 & 12.13 & 8.74 $\pm$ 0.17 & 10.87 \\
GN\_IRS5  &  12:36:20.94 & +62:07:14.0 & 1.15 & 12.23 & 8.64  $\pm$ 0.24 & 10.79 \\
GN\_IRS10 & 12:37:07.21 & +62:14:08.1 & 2.49 & 12.72 & 8.85 $\pm$ 0.12 & 11.07\\
GN\_IRS11 & 12:36:21.27 & +62:17:08.0 & 2.00 & 12.56 (12.55) & 8.84 $\pm$ 0.14 & 11.20 \\
GN\_IRS15 & 12:37:11.37 & +62:13:31.0 & 2.00 & 12.86 (12.85) & 9.20 $\pm$ 0.06 & 11.28\\
GN\_IRS18 & 12:37:16.59 & +62:16:43.0 & 1.80 & 12.46 (12.43) & 8.67 $\pm$ 0.21 & 11.52 \\
GN\_IRS21 & 12:36:18.33 & +62:15:50.0 & 2.00 & 12.77 (12.71) & 8.88 $\pm$ 0.13 & 10.75\\
GN\_IRS23 & 12:36:19.13 & +62:10:04.0 & 2.21 & 12.30 & 8.44 $\pm$ 0.33 & 11.02\\
GN\_IRS25& 12:37:01.59 & +62:11:46.0 & 1.72 & 12.62 & 8.79 $\pm$ 0.16 & 11.54\\
GN\_IRS26 & 12:36:34.51 & +62:12:40.0 & 1.22 & 12.62 & 8.45 $\pm$ 0.36 & 10.94\\
GN\_IRS27 & 12:36:55.94 & +62:08:08.0 & 0.79 & 12.06 (12.03) & 8.37 $\pm$ 0.37 & 11.10\\
GN\_IRS31 & 12:36:22.66 & +62:16:29.0 & 1.79 & 12.44 & 8.54 $\pm$ 0.29 & 11.15\\
GN\_IRS38 &  12:36:29.10 & +62:10:46.3 & 1.01 & 12.00 & 8.75  $\pm$ 0.16 & 11.14\\
GN\_IRS42 & 12:36:46.72 & +62:08:33.9 & 0.97 & 12.13 (12.12) & 8.75  $\pm$ 0.16 & 11.05\\
GN\_IRS45 & 12:38:21.76 & +62:17:06.0 & 1.62 & 12.45 (12.41) & 9.02  $\pm$ 0.19 & 11.09\\
GN\_IRS47 & 12:35:53.81 & +62:13:38.0 & 0.88 & 11.87 (12.86) & 9.01  $\pm$ 0.09 & 10.88\\
GN\_IRS49 &  12:37:05.49 & +62:21:24.0 & 0.95 & 11.85 & 8.62  $\pm$ 0.24 & 11.11\\
GN\_IRS60  & 12:36:49.72 & +62:13:12.9 & 0.47 & 11.24 & 8.44  $\pm$ 0.20 & \, 9.99\\
GN\_IRS62  & 12:36:29.54 & +62:06:46.5 & 0.80 & 11.58 & 8.48  $\pm$ 0.31 & 10.56\\
GN\_IRS66 & 12:37:05.85 & +62:21:29.8 & 0.95 & 11.82 & 8.65  $\pm$ 0.23 & 10.47\\
GN\_IRS67  & 12:36:19.14 & +62:13:01.8 & 1.23 & 11.76 & 8.55  $\pm$ 0.28 & 10.47\\
GN\_IRS68  & 12:36:19.50 & +62:12:52.6 & 0.47 & 11.47 & 8.25  $\pm$ 0.31 & 10.56\\
GS\_IRS20  & 03:32:47.58 & $-$27:44:52.0 & 1.91 & 12.60 (12.59) & 8.55  $\pm$ 0.24 & 10.77\\
GS\_IRS23  & 03:32:17.23 & $-$27:50:37.0 & 1.96 & 12.35 & 8.78  $\pm$ 0.14 & 10.99\\
GS\_IRS25  & 03:32:18.70 & $-$27:49:19.0 & 1.04 & 11.83 & 8.44  $\pm$ 0.28 & 10.60 \\
GS\_IRS26  & 03:32:20.70 & $-$27:44:53.0 & 0.97 & 11.52 & 8.38  $\pm$ 0.31 & 10.69 \\
GS\_IRS28  & 03:31:35.26 & $-$27:49:58.0 & 0.96 & 11.61 & 8.52  $\pm$ 0.23 & 10.62\\
GS\_IRS29  & 03:32:22.53 & $-$27:45:38.0 & 2.08 & 12.12 (12.10) & 8.47  $\pm$ 0.27 & 10.62\\
GS\_IRS30  & 03:32:43.78 & $-$27:52:31.0 & 1.62 & 11.85 & 8.33  $\pm$ 0.38 & 11.07\\
GS\_IRS31  & 03:32:06.00 & $-$27:45:07.0 & 1.07 & 11.78 & 8.36  $\pm$ 0.34 & 10.57\\
GS\_IRS34  & 03:32:39.00 & $-$27:44:20.0 & 1.93 & 11.95 & 8.39  $\pm$ 0.33 & 10.11\\
GS\_IRS39  & 03:31:48.18 & $-$27:45:35.0 & 1.72 & 12.34 (12.31) & 8.79  $\pm$ 0.13 & \nodata\\
GS\_IRS43  & 03:31:48.93 & $-$27:39:45.0 & 0.82 & 11.62 (11.58) & 8.57  $\pm$ 0.19 & 10.34\\
GS\_IRS45  & 03:32:17.45 & $-$27:50:03.0 & 1.62 & 12.48 & 8.64  $\pm$ 0.19 & 10.39\\
GS\_IRS46  & 03:32:42.71 & $-$27:39:27.0 & 1.85 & 12.32 & 8.70  $\pm$ 0.16 & \nodata\\
GS\_IRS52  & 03:32:12.52 & $-$27:43:06.0 & 1.79 & 12.11 & 8.50  $\pm$ 0.25 & 10.43\\
GS\_IRS53  & 03:32:12.13 & $-$27:42:49.0 & 2.45 & 12.37 (12.31) & 8.74  $\pm$ 0.14 & 10.90\\
GS\_IRS56  & 03:32:13.58 & $-$27:47:54.0 & 1.72 & 11.91 & 8.41  $\pm$ 0.31 & 11.22\\
GS\_IRS62  & 03:32:22.48 & $-$27:49:35.0 & 0.73 & 11.76 & 8.56  $\pm$ 0.17 & 10.76\\
GS\_IRS63  & 03:32:22.59 & $-$27:44:25.9 & 0.74 & 11.64 & 8.42  $\pm$ 0.24 & 10.63\\
GS\_IRS64  & 03:32:17.28 & $-$27:49:08.0 & 1.62 & 11.88 & 8.50  $\pm$ 0.26 & 10.45\\
GS\_IRS70  & 03:32:27.71 & $-$27:50:40.6 & 1.10 & 11.99 (11.98) & 8.41  $\pm$ 0.31 & 10.78\\
GS\_IRS73  & 03:32:43.24 & $-$27:47:56.2 & 0.67 & 11.26 & 8.17  $\pm$ 0.41 & 10.47\\
GS\_IRS74  & 03:32:44.32 & $-$27:49:11.9 & 2.00 & 12.12 & 8.56  $\pm$ 0.22 & \nodata\\
GS\_IRS76  & 03:32:48.83 & $-$27:42:35.0 & 1.96 & 12.58 (12.57) & 9.16  $\pm$ 0.06 & \nodata\\
GS\_IRS80  & 03:32:36.52 & $-$27:46:30.7 & 1.05 & 11.91 (11.89) & 8.69  $\pm$ 0.16 & \nodata\\
MIPS289  & 17:13:50.02 & +58:56:57.1 & 1.89 & 13.10 (13.07) & 9.00  $\pm$ 0.07 & \nodata\\
MIPS8377  & 17:17:33.53 & +59:46:40.4 & 0.84 & 12.05 & 8.44  $\pm$ 0.34 & \nodata\\
MIPS8493  & 17:18:05.06 & +60:08:32.6 & 1.80 & 12.73 (12.70) & 8.62  $\pm$ 0.24 & \nodata\\
MIPS8543  & 17:18:12.54 & +59:39:23.0 & 0.65 & 12.03 & 9.06  $\pm$ 0.14 & \nodata\\
MIPS22530  & 17:23:03.33 & +59:16:00.1 & 1.96 & 13.05 (13.04) & 8.97  $\pm$ 0.12 & \nodata\\
19456000  & 17:14:29.66 & +59:32:33.7 & 1.97 & 13.27 (13.26) & 8.93  $\pm$ 0.07 & \nodata
\enddata
\tablenotetext{a}{$L_{\rm IR}^{\rm \scriptscriptstyle SF}$ is listed in parenthesis for those sources requiring a correction to account for nuclear activity as indicated by the presence of a hot dust continuum in the mid-IR spectrum.}
\tablenotetext{b}{$M_\ast$ was initially derived using a Salpeter IMF \citep[see][for details]{kirkpatrick2012}, and we have corrected it to a Chabrier IMF, which we list here.}
\end{deluxetable*}

We use $L_{\rm IR}$ for each sample compiled from the literature (see Section \ref{sec:sample}). In all cases, the $L_{\rm IR}$ was {\it not} determined using the submm data that goes into the $M_{\rm dust}$ calculation, so in this way, $L_{\rm IR}$ and dust mass are independent. $L_{\rm IR}$ for the 5MUSES galaxies and Supersample were measured by fitting template libraries, so the main source of concern is how the choice of $\beta$ used in constructing the templates might effect $L_{\rm IR}$, as $M_{\rm dust}$ also depends on $\beta$. We calculate that varying $\beta\in[1.5,2.0]$ changes $L_{\rm IR}$ by 5\%, so using different values of $\beta$ to calculate $L_{\rm IR}$ and $M_{\rm dust}$ is not a significant source of scatter in the resulting relationships. In the analysis below, we correct $L_{\rm IR}$ for any contribution from nuclear activity as indicated by the presence of a hot dust continuum in the mid-IR spectrum, so that we are actually comparing $M_{\rm dust}$ with $L_{\rm IR}^{\rm \scriptscriptstyle SF}$. The correction is at most 0.07\,dex.
As all of the GOALS and 5MUSES photometry and redshifts used in this paper are published elsewhere in the literature \citep{wu2010,magdis2013,u2012}, we only list the $M_{\rm dust}$ and $L_{\rm IR}$ values that we have calculated for the Supersample in Table \ref{properties}. 

\subsection{Stellar Mass}
Stellar masses are compiled from the literature. For GOALS, \citet{u2012} calculates $M_\ast$ by fitting \citet{bruzual2003} stellar population models to 10 different broadband UV-NIR photometry points (from {\it GALEX}, ground-based observatories, and {\it Spitzer}). The star formation history is parameterized as SFH$=e^{-t/\tau}$ with $\tau\in[1,30]\,$Gyr. Metallicity is a free parameter. The authors choose a Chabrier initial mass function (IMF) \citep{chabrier2003} with a mass range of $0.1\,M_\odot$ to $100\,M_\odot$.

\citet{shi2011} calculate stellar masses for the 5MUSES sample by fitting \citet{bruzual2003} stellar population models to UV-IR data assuming a Chabrier IMF. The SFH has $\tau\in[0.03,22.4]\,$Gyr, and metallicity is a free parameter.

For the GOODS and ECDFS sources (we do not have $M_\ast$ for the six xFLS sources included in the final high $z$ sample), \citet{kirkpatrick2012} describes the procedure for calculating $M_\ast$ by fitting \citet{bruzual2003} models to suite of photometry from the {\it U} band to 4.5\,$\mu$m \citep[see also][]{pannella2015}. For the SFH, $\tau\in[0.1,20]\,$Gyr, and metallicity is fixed to solar. For the Supersample, a Salpeter IMF is used \citep{salpeter1955}, which results in higher stellar masses than a Chabrier IMF. We convert the Supersample masses to a Chabrier framework using 
$M_\ast^{\rm Cha} = 0.62\,M_\ast^{\rm Sal}$ \citep{zahid2012,speagle2014}.

We are drawing from three different samples fit with three different methods, so we verify the stellar masses by comparing with rest frame K-band estimated $M_\ast$ \citep{howell2010}. In all cases, the stellar masses are consistent to within 0.6\,dex, which is the largest deviation. We further check the consistency of the Supersample masses, since these were initially calculated with a Salpeter IMF and require a conversion. We fit the UV-submm SEDs of 10 randomly selected galaxies with the {\sc magphys} code, which employs an energy-balance technique to simultaneously fit the UV/optical/IR emission \citep{dacunha2008}. The {\sc magphys} derived stellar masses are on average 0.2\,dex larger than the masses from \citet{kirkpatrick2012}, converted to a Chabrier IMF. 

All three samples cover a similar range of $M_\ast = 1\times 10^{10}-3\times10^{11}\,M_\odot$, as can be seen in Figure \ref{stellar_mass}.

\section{SED Comparison}
\label{sec:shape}
\begin{figure}
\centering
\includegraphics[width=3.2in]{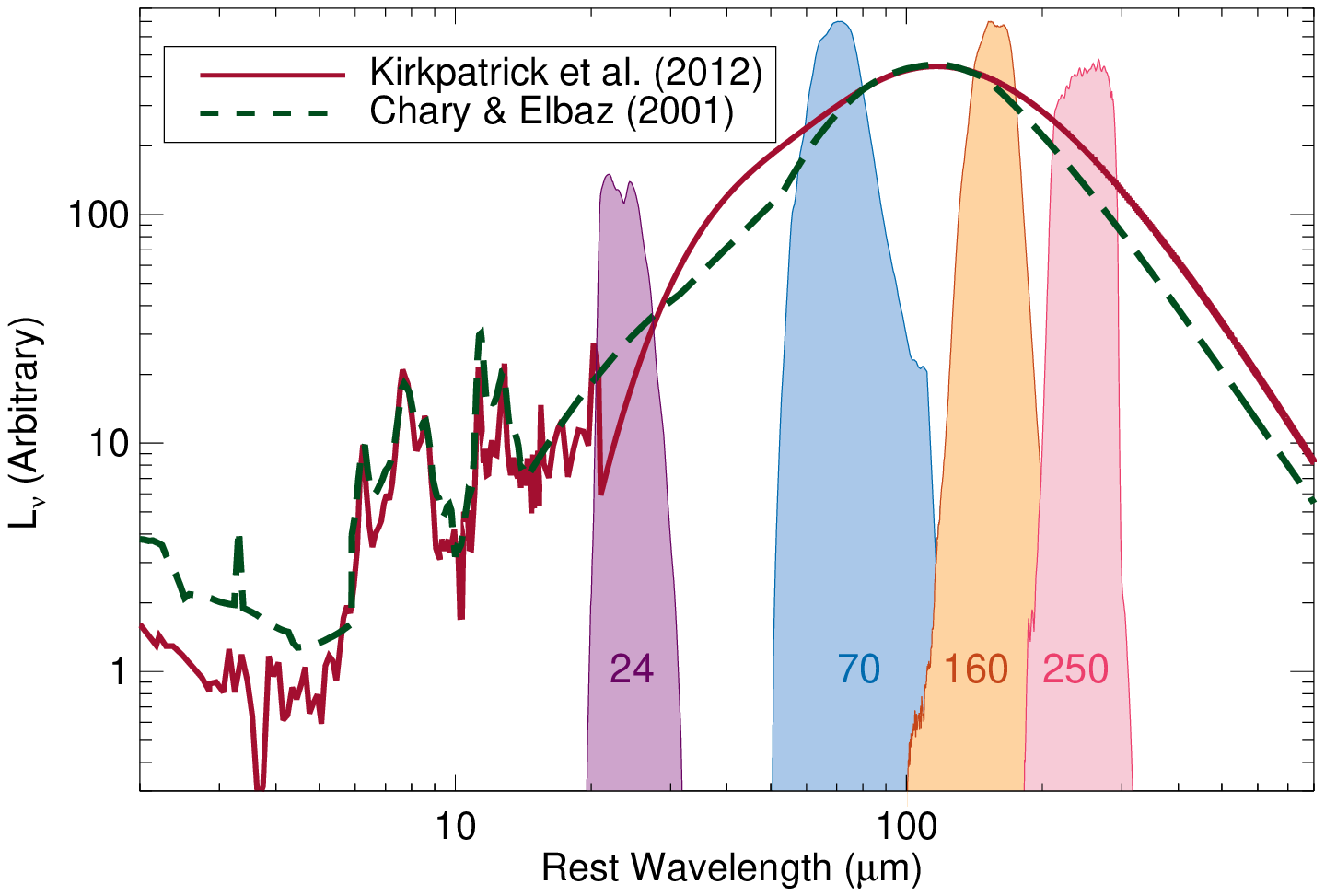}
\caption{We illustrate where the broadband MIPS 24, 70, 160\,$\mu$m and SPIRE 250\,$\mu$m transmission filters sample the SED using a representative SFG template (red line) at $z\sim1$ \citep{kirkpatrick2012} with $\log L_{\rm IR}=11.62\,L_\odot$. To illustrate how the far-IR SED changes, we also plot a template (green dashed line) from the local \citet{chary2001} library, with $\log L_{\rm IR} = 10.79\,L_\odot$. The templates have been arbitrarily normalized so that the peaks coincide. The $z\sim1$ template has proportionally more cold dust, notably at 250\,$\mu$m, than the local template.\label{template_illustration}}
\end{figure}

We look for evolution in the IR SED of our samples of DSFGs by comparing rest frame colors, as this does not require any model assumptions.
We show a sample SED at $z\sim1$ and the MIPS and SPIRE 250\,$\mu$m broadband photometry filters in Figure \ref{template_illustration} as a visual guide when comparing galaxy colors below. We also plot a
local template from the \citep{chary2001} library to better illustrate how the far-IR colors can indicate SED shape. We plot a local template that peaks at the same wavelength as our $z\sim1$ template, despite being nearly an order of magnitude lower in luminosity, so that the color $L_{160}/L_{70}$ should be approximately equal for both the local and $z\sim1$ template. However, the $z\sim1$ template has proportionally more cold dust, as evidenced by an increase in the emission longwards of $\sim200\,\mu$m.

\begin{figure}
\centering
\includegraphics[width=3.3in]{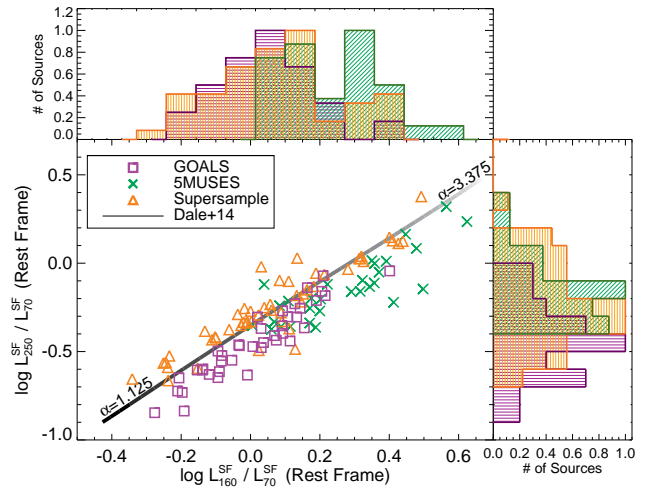}
\caption{$L_{160}^{\rm \scriptscriptstyle SF}/L_{70}^{\rm \scriptscriptstyle SF}$, a proxy for $T_{\rm dust}$, versus $L_{250}^{\rm \scriptscriptstyle SF}/L_{70}^{\rm \scriptscriptstyle SF}$, which measures the width of the SED. These colors are tightly correlated and increase with decreasing strength of the incident radiation field, as parameterized by the \citet{dale2014} library (thick black/grey line). There is an offset between the low $z$ galaxies (purple squares and green crosses) and the high $z$ galaxies (orange triangles) , with the high $z$ sample having higher $L_{250}^{\rm \scriptscriptstyle SF}/L_{70}^{\rm \scriptscriptstyle SF}$ for a given $L_{160}^{\rm \scriptscriptstyle SF}/L_{70}^{\rm \scriptscriptstyle SF}$, indicating enhanced submm emission and possibly higher dust masses. \label{s250_70_160}}
\end{figure}

In Figure \ref{s250_70_160}, we compare $L_{160}^{\rm \scriptscriptstyle SF}/L_{70}^{\rm \scriptscriptstyle SF}$ and $L_{250}^{\rm \scriptscriptstyle SF}/L_{70}^{\rm \scriptscriptstyle SF}$. The peak of the SED is traced by $L_{160}^{\rm \scriptscriptstyle SF}/L_{70}^{\rm \scriptscriptstyle SF}$, which is also a proxy for $T_{\rm dust}$. $L_{250}^{\rm \scriptscriptstyle SF}/L_{70}^{\rm \scriptscriptstyle SF}$ is difficult to interpret decoupled from $L_{160}^{\rm \scriptscriptstyle SF}/L_{70}^{\rm \scriptscriptstyle SF}$. If $L_{160}^{\rm \scriptscriptstyle SF}/L_{70}^{\rm \scriptscriptstyle SF}$ indicates temperature, then for a given $T_{\rm dust}$, $L_{250}^{\rm \scriptscriptstyle SF}/L_{70}^{\rm \scriptscriptstyle SF}$ can be thought of as the amount of cold dust (traced by $L_{250}^{\rm \scriptscriptstyle SF}$) relative to warm dust (traced by $L_{70}^{\rm \scriptscriptstyle SF}$). SEDs with proportionally more warm dust will have smaller ratios of $L_{250}^{\rm \scriptscriptstyle SF}/L_{70}^{\rm \scriptscriptstyle SF}$. All three samples have a tight relationship between $L_{160}^{\rm \scriptscriptstyle SF}/L_{70}^{\rm \scriptscriptstyle SF}$ and $L_{250}^{\rm \scriptscriptstyle SF}/L_{70}^{\rm \scriptscriptstyle SF}$. 

Physically, this relationship can be described by how the incident radiation field in the ISM drives the far-IR emission \citep{witt2000,misselt2001,dale2002,dale2014}, which we illustrate using the suite of IR star forming templates from \citet{dale2014}. These model SEDs are created by combining the emission from different dust mass elements, heated by an incident radiation field of strength $U$, in a power law distribution:
\begin{equation}
\label{eq:rad}
d M_{\rm dust} \propto U^{-\alpha} d U
\end{equation}
We chose the \citet{dale2014} templates due to the simplicity of interpretation. In this formulation, $\alpha$ characterizes the relative contributions of each dust mass element to the total SED, so that a smaller $\alpha$ means more of the dust in a galaxy is heated by a harsher radiation field.
We show the location of the \citet{dale2014} templates in Figure \ref{s250_70_160} using the thick grey-scale line. Our sources cluster around this line and have colors suggestive of
$\alpha\in[1.125,3.375]$, with the low $z$ (GOALS + 5MUSES) and high $z$ samples spanning a similar range of $\alpha$. 

The two low redshift samples show a distinct offset along the dark line, which is not unexpected given the different luminosities probed by 5MUSES and GOALS. We measure this offset to be 0.2\,dex by fitting a line to the high $z$ and low $z$ (GOALS + 5MUSES) samples separately. %The GOALS sources are more luminous, indicative of a stronger radiation field, and therefore they have a lower $L_{160}/L_{70}$ (warmer dust temperature).
Supersample sources in general have a higher $L_{250}^{\rm \scriptscriptstyle SF}/L_{70}^{\rm \scriptscriptstyle SF}$ than GOALS or 5MUSES sources at fixed $L_{\rm 160}^{\rm \scriptscriptstyle SF}/L_{\rm 70}^{\rm \scriptscriptstyle SF}$ (or $T_{\rm dust}$), although both the high $z$ and low $z$ samples span the same range in each individual color. This suggests an evolution with redshift of the longer wavelength dust emission, due either to increased dust mass or possibly a colder dust component boosting the submm emission \citep[e.g.,][]{galametz2014}. %indicating proportionally more cold dust emission with increasing redshift for (U)LIRGs. 

\subsection{Trends with $L_{\rm IR}^{\rm \scriptscriptstyle SF}$}
\begin{figure}
\centering
\includegraphics[width=3.3in]{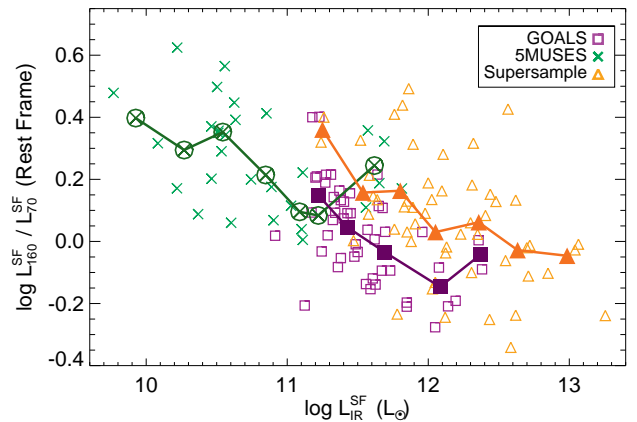}
\caption{$L_{160}^{\rm \scriptscriptstyle SF}/L_{70}^{\rm \scriptscriptstyle SF}$ vs. $L_{\rm IR}^{\rm \scriptscriptstyle SF}$ for the GOALS and Supersample. We overplot the means as the filled symbols. This correlation is stronger for the low $z$ samples (which have a Kendall's $\tau$ of $-0.51$), illustrating the increasing diversity of of DSFGs at high $z$ (a Kendall's $\tau$ test shows $\tau= -0.33$ for the Supersample. There is a clear offset between the samples, with the Supersample having redder colors (colder $T_{\rm dust}$), due to the evolution of the $L_{\rm IR} -T$ relationship with redshift. \label{color_lir}}
\end{figure}

Our higher $z$ sample is also at a higher $L_{\rm IR}$, which can confuse any redshift evolution. We compare $L_{160}^{\rm \scriptscriptstyle SF}/L_{70}^{\rm \scriptscriptstyle SF}$ with $L_{\rm IR}^{\rm \scriptscriptstyle SF}$ in Figure \ref{color_lir}. Figure \ref{color_lir} shows a clear trend between $L_{160}^{\rm \scriptscriptstyle SF}/L_{70}^{\rm \scriptscriptstyle SF}$ and $L_{\rm IR}^{\rm \scriptscriptstyle SF}$, otherwise known as the $L_{\rm IR}-T_{\rm dust}$ relation \citep{chapman2003,casey2012,magnelli2014,lee2016}. We overplot the mean of each sample in $\log(L_{\rm IR}^{\rm \scriptscriptstyle SF})$ bins of 0.25 as the larger symbols. 

The 5MUSES and GOALS galaxies follow the same general trend, but there is a clear offset between the low $z$ samples and Supersample, due to evolution of the $L_{\rm IR}-T_{\rm dust}$ relationship with redshift \citep{chapin2009,casey2012,magnelli2014}. 
 At a given $L_{\rm IR}^{\rm \scriptscriptstyle SF}$, the difference between low $z$ and high $z$ is $\sim0.2$ dex, which corresponds to approximately a 5\,K temperature difference. The Supersample galaxies have a much larger scatter than the GOALS sample, illustrating the increasing diversity of DSFGs with redshift \citep{symeonidis2010}. Interestingly, the 5MUSES galaxies with $\log L_{\rm IR}^{\rm \scriptscriptstyle SF}>11.4$ overlap with the Supersample rather than the GOALS sample. These are also the highest redshift 5MUSES galaxies, with $z>0.15$, and they also overlap with the Supersample below in Figure \ref{mdust_lir}.
Overall, we see a similar slope in the mean $L_{160}^{\rm \scriptscriptstyle SF}/L_{70}^{\rm \scriptscriptstyle SF}$ vs. $L_{\rm IR}^{\rm \scriptscriptstyle SF}$ points, with a normalization that depends on redshift, in agreement with the findings in \citet{magnelli2014} and \citet{casey2012}. 

\section{Evolving Trends with Dust Mass}
\label{sec:ism}
\begin{figure}
\includegraphics[width=3.3in]{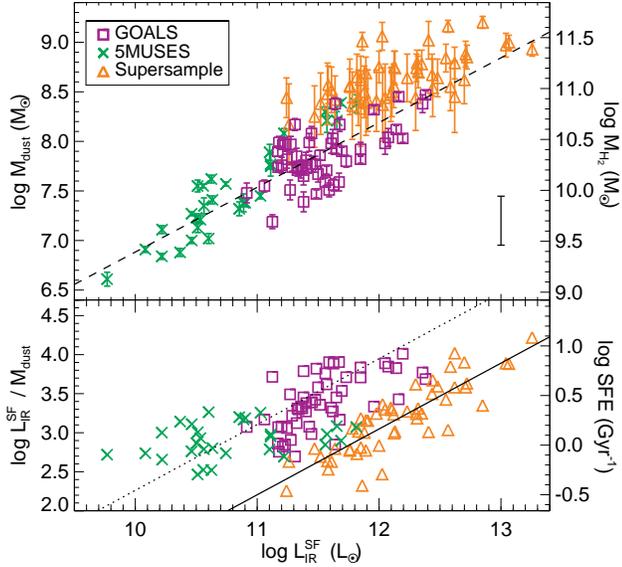}
\caption{{\it Top--}$M_{\rm dust}$ v.\,$L_{\rm IR}^{\rm \scriptscriptstyle SF}$ for $T_{\rm cold}=25\,$K. There is a tight correlation, which we parameterize with the dashed line, fit to the GOALS (purple squares) and 5MUSES (green crosses) sources. The Supersample (orange triangles) lies above this relation. %We include normal star forming galaxies from the KINGFISH survey (grey crosses) for comparison. 
In this and subsequent Figures \ref{mdust_lir_color} and \ref{Tdiff}, the uncertainties on $M_{\rm dust}$ simply reflect the photometric uncertainties on $S_\nu$. The effect of changing $T_{\rm cold}$ by $\pm5\,$K is illustrated by the solid bar in the bottom right corner. {\it Bottom--} $\log L_{\rm IR}^{\rm \scriptscriptstyle SF}/M_{\rm dust}$ as a function of $L_{\rm IR}^{\rm \scriptscriptstyle SF}$. The evolution with redshift is clearly seen, as the Supersample lies distinctly offset from the 5MUSES and GOALS samples. \label{mdust_lir}}
\end{figure}

We now consider the physical parameter, $M_{\rm dust}$. We have calculated $M_{\rm dust}$ in a self-consistent manner for all galaxies, while making as few assumptions as possible.
Figure \ref{mdust_lir} shows a clear, strong trend between $M_{\rm dust}$ and $L_{\rm IR}^{\rm \scriptscriptstyle SF}$ (Kendall's $\tau = 0.68$). The dashed line is fit to the 5MUSES and GOALS galaxies, and almost all the Supersample sources lie above this relation. The fit is sub linear, with
\begin{equation}
\label{eq:ML}
M_{\rm dust} = 2.26 \times (L_{\rm IR}^{\rm \scriptscriptstyle SF})^{0.653}
\end{equation}
For clarity, the error bars on each point only reflect the error of the submm flux used to calculate $M_{\rm dust}$. 
The effect of changing $T_{\rm cold}$ by $\pm5\,$K is illustrated by the solid bar in the bottom right corner of Figure \ref{mdust_lir} (Han et al.\,in prep). At first approximation, one might expect that $L_{\rm IR}/M_{\rm dust} \propto T_{\rm dust}^{4+\beta}$ in the optically thin regime, if the dust emission can be represented with a single temperature modified blackbody \citep{lanz2014,magnelli2014}. This simplification would imply a linear relationship between $L_{\rm IR}$ and $M_{\rm dust}$ for a fixed $T_{\rm dust}$. Since we measure a sub linear trend, this means that this model is unphysical since dust has a range of temperatures and other effects, such as optical depth or geometry, are affecting the far-IR/submm emission. 

Although in the top panel of Figure \ref{mdust_lir}, it appears that all samples may actually lie along the same relationship between $L_{\rm IR}^{\rm \scriptscriptstyle SF}$ and $M_{\rm dust}$, two separate relationships become evident in the bottom panel of Figure \ref{mdust_lir}, where we show $L_{\rm IR}^{\rm \scriptscriptstyle SF}/M_{\rm dust}$ v.\,$L_{\rm IR}^{\rm \scriptscriptstyle SF}$. There is a strong trend between $L_{\rm IR}^{\rm \scriptscriptstyle SF}/M_{\rm dust}$ and $L_{\rm IR}^{\rm \scriptscriptstyle SF}$, with more luminous galaxies having proportionally less dust for a given radiation field, which results in warmer temperatures (Kendall's $\tau = 0.49\ (0.67)$ for the low (high) $z$ samples). However, this trend (plotted as the solid line) shifts with redshift, so that Supersample galaxies with $L_{\rm IR}^{\rm \scriptscriptstyle SF}=10^{12}\,L_\odot$ have $L_{\rm IR}^{\rm \scriptscriptstyle SF}/M_{\rm dust}$ ratios similar to GOALS galaxies with $L_{\rm IR}^{\rm \scriptscriptstyle SF}=2\times10^{11}\,L_\odot$ (shown by the dotted line); a complementary result was found for submillimeter galaxies in \citet{chakrabarti2008}. The 5MUSES galaxies have a flatter distribution, although the galaxies that clearly overlap the Supersample are the galaxies at higher redshifts ($z>0.15$). Similarly, \citet{magdis2014} found that intermediate redshift DSFGs had larger gas reservoirs and lower dust temperatures than their local counterparts by a redshift of $z\sim0.33$. %If there is a change in $L_{\rm IR}/M_{\rm dust}$ with redshift, then 

Submm fluxes can also be used to calculate gas mass \citep{scoville2015}, although this requires extra assumptions, such as the expected dust-to-gas ratio. We illustrate on the right axis in Figure \ref{mdust_lir} what the H$_2$ gas mass would be, following the formalism of \citet{scoville2015}:
\begin{equation}
M_{\rm H_2} = \alpha_{850}^{-1} L_{850}
\end{equation}
where $\alpha_{850}=6.7\times10^{19}$\,ergs\,s$^{-1}$\,Hz$^{-1}$\,M$_\odot^{-1}$. We have CO luminosities for a handful of 5MUSES and Supersample galaxies \citep{kirkpatrick2014,yan2010}, and $M_{\rm H_2}$ calculated from $L_{\rm CO}$ for these galaxies agrees with $M_{\rm H_2}$ calculated from $L_{\rm 850}$. Furthermore, the gas masses we derive here for our high $z$ sample are completely consistent with the gas masses in \citet{scoville2015} for galaxies spanning the same $L_{\rm IR}$ and redshift ranges. We then convert $L_{\rm IR}^{\rm \scriptscriptstyle SF}$ to SFR with ${\rm SFR}\,[M_\odot\,{\rm yr}^{-1}] = 1.59\times 10^{-10} \times {L_{\rm IR}^{\rm \scriptscriptstyle SF}\,[L_\odot]}$ from \citet{murphy2011}. In the bottom panel, we show how $L_{\rm IR}^{\rm \scriptscriptstyle SF}/M_{\rm dust}$ translates to SFR$/M_{\rm H_2}$, which is the star formation efficiency (SFE). The relationship between SFE and $L_{\rm IR}^{\rm \scriptscriptstyle SF}$ evolves with redshift. That is, at a given $L_{\rm IR}^{\rm \scriptscriptstyle SF}$, local galaxies convert gas to stars {\it more} efficiently than high $z$ galaxies. Assuming a dust-to-gas ratio of $\sim$100, the $L_{\rm IR}^{\rm \scriptscriptstyle SF}/M_{\rm dust}$ relation for the Supersample is in excellent agreement with that predicted by hyrdodynamical simulations combined with radiative transfer \citep[Fig. 9 of][]{hayward2012}. We remind the reader that we are comparing galaxies at a fixed luminosity, rather than a fixed location on the main sequence, which we examine in Section \ref{sec:gas}. At $L_{\rm IR}^{\rm \scriptscriptstyle SF}>10^{11}$, local DSFGs lie predominantly above the main sequence, while $z\sim1-2$ DSFGs lie on it, which would account for the difference in SFEs. 
 
\begin{figure}
\includegraphics[width=3.3in]{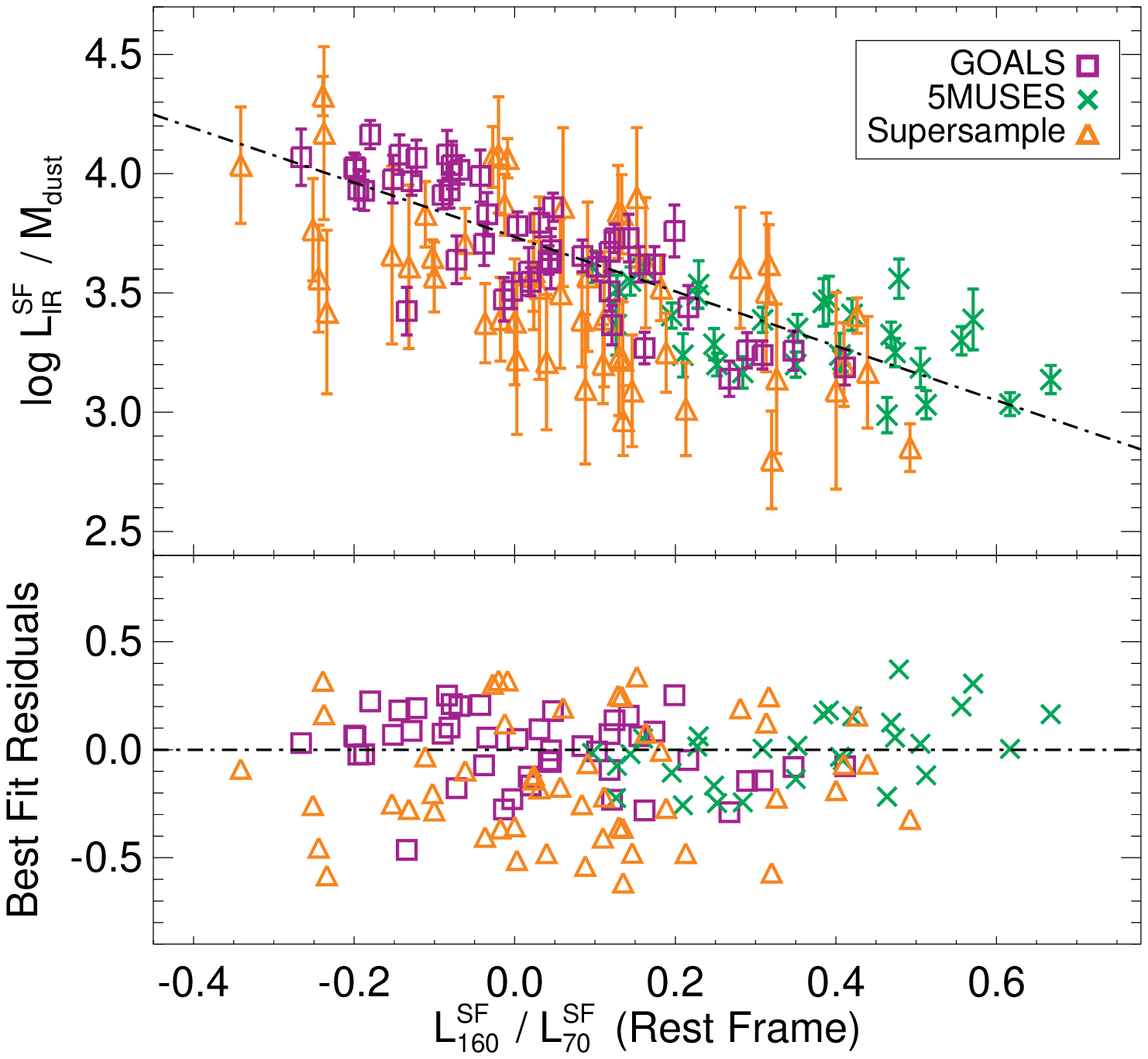}
\caption{{\it Top--} $L_{\rm IR}^{\rm \scriptscriptstyle SF}/M_{\rm dust}$ vs. $\log L_{160}^{\rm \scriptscriptstyle SF}/L_{70}^{\rm \scriptscriptstyle SF}$. There is a strong correlation between the two parameters, as exhibited by the dot dashed line, which is the best fit to all the data (Equation \ref{eq:MLC}). The relationship between $L_{\rm IR}^{\rm \scriptscriptstyle SF}/ M_{\rm dust}$ and $L_{160}^{\rm \scriptscriptstyle SF}/L_{70}^{\rm \scriptscriptstyle SF}$ does not evolve with redshift, and so $L_{\rm IR}^{\rm \scriptscriptstyle SF}/M_{\rm dust}$ alone can account for the colder temperatures in high $z$ galaxies. {\it Bottom--} Residuals around the best fit line shown in the top panel. On average the Supersample has a mean residual of -0.14\,dex, indicating they lie slightly below the low $z$ samples. However, the scatter is quite large, with a standard deviation of 0.28\,dex, arguing against any clear redshift evolution.\label{mdust_lir_color}}
\end{figure}

We now return to the question of what drives the cooler dust temperatures observed in higher redshift galaxies.
Based on simulations, \citet{saf2015} predict that $L_{\rm IR}$ and $M_{\rm dust}$ alone determine the shape of the SED, independent of the size of the galaxy. That is, $T_{\rm dust}$ depends solely on the total luminosity absorbed by dust ($L_{\rm IR}$) and the amount of absorbing material ($M_{\rm dust}$), in which case it is not necessary to measure all three parameters for a galaxy, since the third parameter can be inferred from the other two. We test this prediction with observations in Figure \ref{mdust_lir_color}, where we plot $L_{\rm IR}^{\rm \scriptscriptstyle SF}/M_{\rm dust}$ vs. $L_{160}^{\rm \scriptscriptstyle SF}/L_{70}^{\rm \scriptscriptstyle SF}$, a proxy for $T_{\rm dust}$. 

There is a clear correlation between the parameters (Kendall's $\tau = -0.51$), and crucially, there is no offset with redshift. \citet{magdis2012} found that $L_{\rm IR}/M_{\rm dust}$ is proportional to $\langle U\rangle$, the average interstellar radiation field. In turn, $\langle U \rangle$ determines $T_{\rm dust}$, giving rise to the strong correlation \citep{draine2007}. The dot-dashed line shows the relationship:
\begin{equation}
\label{eq:MLC}
\log \left(\frac{L_{\rm IR}^{\rm \scriptscriptstyle SF}}{M_{\rm dust}}\right)=  -1.24 \times \log \left(\frac{L_{160}^{\rm \scriptscriptstyle SF}}{L_{70}^{\rm \scriptscriptstyle SF}}\right) + 3.68
\end{equation}
which we have derived using all three samples, so this parameterization holds over three orders of magnitude ($L_{\rm IR}^{\rm \scriptscriptstyle SF}=10^{10}-10^{13}\,L_\odot$) and from $z\sim0-2$. This parameterization can be used to estimate $M_{\rm dust}$ from far-IR data without requiring a longer wavelength ($\lambda > 500\,\mu$m) observation. This relationship is also observed on much smaller scales ($\leq1\,$kpc$^2$) in local resolved galaxies from the KINGFISH survey \citep{kirkpatrick2014a}. We calculate the residuals around the best fit line and show these in the bottom panel of Figure \ref{mdust_lir_color}. Although the Supersample have a mean residual of -0.14\,dex, indicating they lie slightly below the best fit line, the scatter of the residuals is quite large (standard deviation of 0.28\,dex). Combined with the large uncertainties on the dust masses, this argues against any strong redshift evolution in Equation \ref{eq:MLC}.

With the data available to us, the source of scatter in this Figure is open to interpretation. 
%The derivation of the dust masses are the most uncertain, whereas $L_{\rm IR}^{\rm \scriptscriptstyle SF}$ is a fairly robust parameter. An evolving temperature in the cold ISM could produce the scatter in $M_{\rm dust}$, since we have held this constant for all galaxies. In this way, galaxies at a fixed $L_{160}^{\rm \scriptscriptstyle SF}/L_{70}^{\rm \scriptscriptstyle SF}$ should have the same $L_{\rm IR}^{\rm \scriptscriptstyle SF}/M_{\rm dust}$, and we should change $T_{\rm dust}$ and $\kappa_\nu$ accordingly when measuring $M_{\rm dust}$.
One likely source is the geometry of the ISM, which can explain how galaxies with the same $L_{\rm IR}^{\rm \scriptscriptstyle SF}/M_{\rm dust}$ can have different $L_{160}^{\rm \scriptscriptstyle SF}/L_{70}^{\rm \scriptscriptstyle SF}$ ratios. A galaxy whose ISM approximates a shell geometry, where the stars are enclosed by a spherical shell of dust, will always peak at shorter wavelengths than an ISM geometry where the stars and dust are well mixed, all other parameters being equal \citep{misselt2001}. This is because the shell geometry absorbs all of the incident radiation from stars, heating the dust to higher temperatures. Testing whether different geometries can account for all of the scatter between $L_{\rm IR}^{\rm \scriptscriptstyle SF}/M_{\rm dust}$ and $L_{160}^{\rm \scriptscriptstyle SF}/L_{70}^{\rm \scriptscriptstyle SF}$ will require resolved observations of the ISMs with ALMA.

The relationship between $L_{\rm IR}^{\rm \scriptscriptstyle SF}/M_{\rm dust}$ and $L_{160}^{\rm \scriptscriptstyle SF}/L_{70}^{\rm \scriptscriptstyle SF}$ can also explain the offset between the low and high $z$ samples in Figure \ref{color_lir}. It is possible that the normalization of the $L_{\rm IR}- T_{\rm dust}$ relation evolves with redshift because galaxies at $z=1-2$ have higher dust masses at a given $L_{\rm IR}$ \citep{magdis2012}. If at fixed $L_{\rm IR}$, the typical dust mass is higher in high redshift DSFGs than their $z\sim0$ counterparts, then the dust temperature will be lower, shifting the normalization of the $L_{\rm IR} - T_{\rm dust}$ relation \citep{saf2015}. This explanation for the shift does not require any change in the mode of star formation or ISM density (compact starburst vs. main sequence) or possibly in the dust composition. To produce colder temperatures for a given radiation field, an increase in the amount of large dust grains, relative to small ones, is required if the far-IR emission is optically thin. However, in the ULIRG regime, the high dust masses can cause the dust to be optically thick even at submm wavelengths \citep[e.g.,][]{alfonso2004}, in which case simply adding more dust (and not changing the proportion of large to small grains) will decrease the temperature.

\section{Discussion}

\subsection{Potential Evolution in Dust Properties?}
\label{sec:SFE}
The DSFGs at $z\sim1-2$ have higher dust masses for a given $L_{\rm IR}^{\rm \scriptscriptstyle SF}$ than local DSFGs. Here we want to make an important distinction. The observed quantity in our sample is the luminosity density in the Rayleigh Jeans tail, and it is the luminosity density that is actually higher in high redshift galaxies (as we have used the same assumptions for all other parameters when calculating $M_{\rm dust}$). The luminosity density in the Rayleigh Jeans tail is directly proportional to $M_{\rm dust} \times \kappa_\nu$, meaning that in principal, $M_{\rm dust}$ and $\kappa_\nu$ are degenerate. Thus far, we have been assuming the same opacity relationship for all galaxies, but in reality, the opacity might be evolving with redshift rather than the $M_{\rm dust}$. Dust opacity has a wavelength dependence
\begin{equation}
\kappa_\nu = \kappa_0 \left(\frac{\lambda}{\lambda_0}\right)^{-\beta}
\end{equation}
so that either the normalization, $\kappa_0$, or the slope, $\beta$, could be evolving.

Although the \citet{weingartner2001} models all have similar $\kappa_\nu$ values in the submillimeter, grain compositions and distributions that differ considerably from these models can give different submillimeter opacities. The apparent increase in dust mass could possibly be due to a change in the dust grain properties with redshift \citep{dwek2014}. For example, ice mantles can affect $\beta$, causing opacity to increase, and ice mantles should be prevalent in cold star forming regions \citep{tielens1984,bergin2000,spoon2002}. The presence of ice mantles can increase $\kappa_\nu$ by a factor of 4 \citep{preibisch1993,pollack1994} %lower $\beta$ from 2 to 1.4, increasing $\kappa$ by a factor of 4 \citep{preibisch1993,pollack1994}. 
A factor of 4 increase in $\kappa$ for the high redshift galaxies would reconcile the difference in $M_{\rm dust}$ for the GOALS and Supersample. The strength of water ice, silicate absorption, and 3.4\,$\mu$m PAH emission features has been measured in a handful of Supersample sources and closely resembles what is measured in local GOALS galaxies with similar obscuration, providing some indication that the amount of ice in dusty galaxies does not strongly evolve with redshift \citep{sajina2009}.

If $\beta$ in our galaxies differs significantly from $\beta$ in the \citet{weingartner2001} models, for example, varying with redshift, then our assumed $\kappa_\nu$ will be incorrect.
The emissivity measure from the Rayleigh Jeans tail is observed to be shallower in lower metallicity galaxies \citep{galametz2011,kirkpatrick2013b}, though there is little evidence that metallicity evolves strongly in massive, dusty galaxies out to $z\sim2$ with similar $M_\ast$, SFR, and $M_{\rm mol}$ \citep{mannucci2010,magdis2012,bothwell2016}. In fact, when we measure $\beta$ directly for individual galaxies by fitting a modified blackbody with $T=25\,$K, we find $\beta=1.8$ for the Supersample and $\beta=2.02$ for the GOALS sample. We are not fully sampling the Rayleigh Jeans tail in the Supersample, and longer wavelength observations are required to more accurately measure the effective $\beta$, but with current observations, we see little evidence supporting a drastic increase in opacity required to reconcile dust mass measurements of our samples. Longer wavelength observations, such as with ALMA or the Large Millimeter Telescope, are required to better measure the effective $\beta$ at $z=1-2$.

\subsection{Potential Role of Mergers or Compactness?}
Additional insight into the ISM of galaxies can be gained by examining morphologies and merger signatures, which we have for the GOALS and Supersample sources. Galaxies hosting a compact central starburst will have a higher temperature than galaxies at the same $L_{\rm IR}$ which are disks \citep{chanial2007}. In local LIRGs and ULIRGs, a compact starburst is triggered by a major merger, as the transfer of angular momentum funnels gas to the central kpc of the merging system. %Advanced mergers should have a more irregular shape, with a central, possibly dust-enshrouded, compact starburst. On the other hand, isolated galaxies are typically disks and are undergoing secular star formation with long gas depletion timescales and low star formation efficiencies. We now examine whether our local and high redshift samples have differing merger fractions or morphologies which could account for the difference in $M_{\rm dust}$, $L_{\rm IR}$, or SFE. 

\citet{stierwalt2013} classified the GOALS sample into merger classifications using {\it HST} $B-$, $I-$, and $H-$band imaging \citep[see also][]{haan2011} and IRAC 3.6\,$\mu$m imaging. All 46  of the galaxies in the present work also have a merger classification. The merger classifications we use here are {\bf 0:} no merger or massive neighbor; {\bf 1:} galaxy pair; {\bf 2:} early-stage (disk still in tact); {\bf 3:} mid-stage (amorphous disk, tidal tails); {\bf 4:} late-stage (two nuclei in common envelope).
Morphology analysis of the Supersample is less straightforward due to our combining different surveys. GOODS-S and GOODS-N galaxies have been visually classified with {\it HST} WFC3 images as part of the Cosmic Assembly Near-ir Deep Extragalactic Legacy Survey (CANDELS; P.I. S. Faber \& H. Ferguson). The CANDELS tool gives classifiers a choice of 5 different interaction stages, similar to those listed above. The morphology catalogs then contain the fraction of classifiers that selected each interaction stage \citep{kartaltepe2015}. We use these fractions to determine a weighted average interaction stage, from $0-4$ \citep[see also][which create a weighted average on a 0-1 scale]{rosario2015}.
\citet{zam2011} classifies 134 xFLS sources into merger classifications using {\it Hubble} NICMOS imaging, from 0-5. Here, we combine old mergers (5) with advanced mergers (4), to match the categories described above. In total, 17 Supersample galaxies have dust masses and a merger classification. 5MUSES galaxies lack a morphology classification.

\begin{figure}
\includegraphics[width=3.3in]{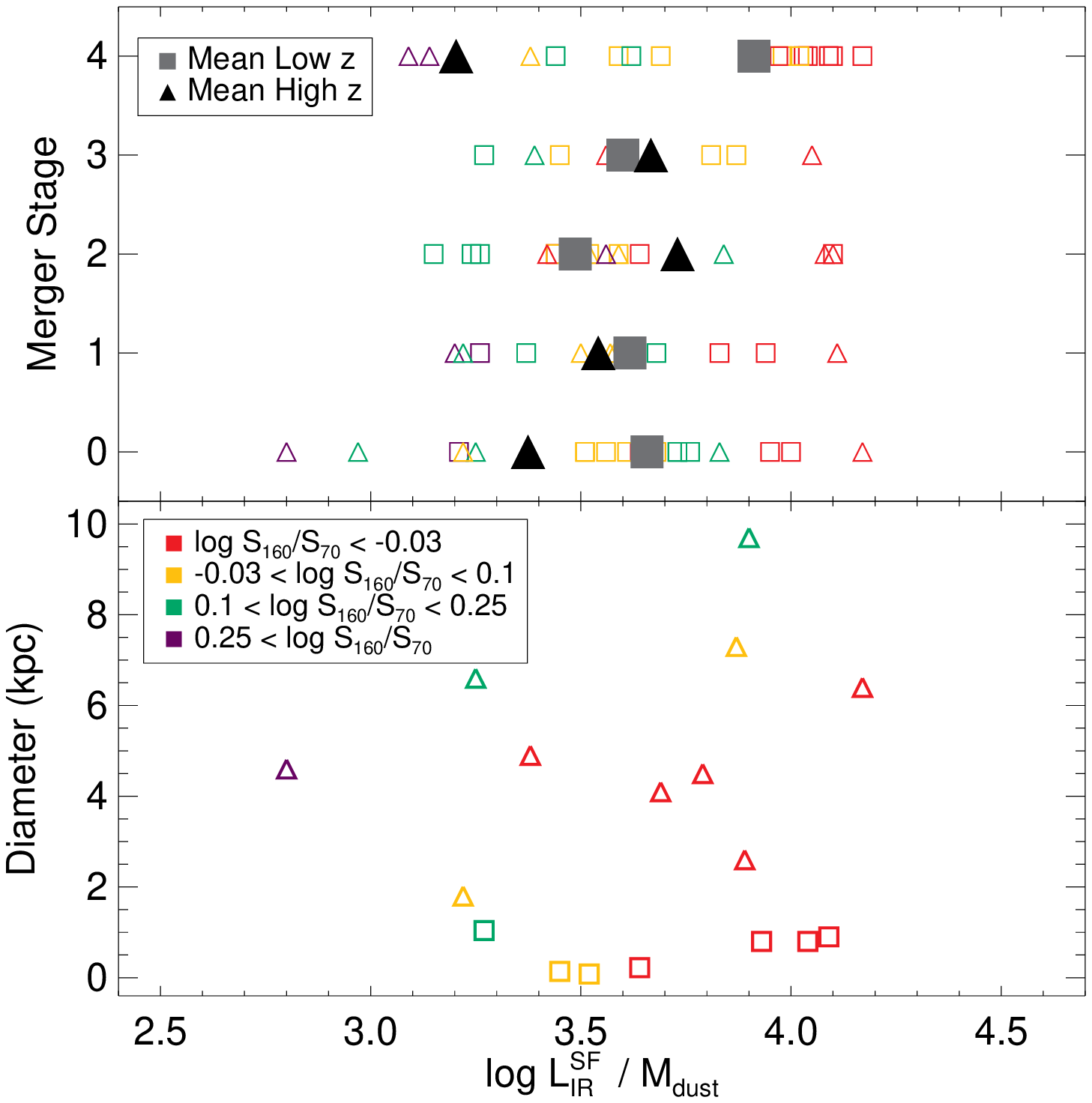}
\caption{{\it Top panel --} merger classification vs. $L_{\rm IR}^{\rm \scriptscriptstyle SF}/M_{\rm dust}$, where sources are shaded by $L_{\rm 160}/L_{70}^{\rm \scriptscriptstyle SF}$. There is no correlation between merger classification and $L_{\rm IR}^{\rm \scriptscriptstyle SF}/M_{\rm dust}$ or $L_{160}^{\rm \scriptscriptstyle SF}/L_{70}^{\rm \scriptscriptstyle SF}$. To help guide the eye, we plot the mean $L_{\rm IR}^{\rm \scriptscriptstyle SF}/M_{\rm dust}$ in each merger stage for the low $z$ galaxies (large grey squares) and high $z$ galaxies (large black triangles). {\it Bottom panel --} ISM diameters from \citet{rujopakarn2011}, measured using CO, radio, or Pa$\alpha$, as a function of $L_{\rm IR}^{\rm \scriptscriptstyle SF}/M_{\rm dust}$. While the high $z$ galaxies are all clearly larger than the GOALS galaxies, there is no correlation between size and $L_{\rm IR}^{\rm \scriptscriptstyle SF}/M_{\rm dust}$. The far-IR/submm data seems to tell observers very little about the compactness or merger stage of galaxies.\label{lir_morph}}
\end{figure}

The top panel of Figure \ref{lir_morph} shows interaction stage as a function of $L_{\rm IR}^{\rm \scriptscriptstyle SF}/M_{\rm dust}$, and the points are shaded by $L_{160}^{\rm \scriptscriptstyle SF}/L_{70}^{\rm \scriptscriptstyle SF}$. There is no correlation between the parameters, for either the GOALS or Supersample. Galaxies at a given $L_{\rm IR}^{\rm \scriptscriptstyle SF}/M_{\rm dust}$ occupy all merger classifications, as do galaxies at a given $L_{160}^{\rm \scriptscriptstyle SF}/L_{70}^{\rm \scriptscriptstyle SF}$. %We remind the reader that we have removed galaxies with hot central sources (AGN or starburst).
DSFGs exhibit a variety of $L_{\rm IR}^{\rm \scriptscriptstyle SF}/M_{\rm dust}$ ratios and a variety of major merger classifications, indicating that the major merger scenario cannot directly explain all of the far-IR observed properties of these galaxies. In the same vein, using hydrodynamical simulations post-processed with dust radiative transfer, \citet{lanz2014} found that the effective dust temperature of mergers does not reflect the merger stage except for a very short period ($\lesssim100$\,Myr) during the coalescence-induced starburst.

The most direct test of how ISM geometry affects $L_{\rm IR}^{\rm \scriptscriptstyle SF}/M_{\rm dust}$ and $T_{\rm dust}$ is to compare the spatial extent of the dusty ISM. Measuring submm sizes on a resolved scale at low and high redshift requires the sensitivity of ALMA. To date, only a handful of DSFGs have resolved ALMA imaging, but we can compare other tracers of the ISM until we have a larger ALMA sample. \citet{rujopakarn2011} compiles from the literature ISM diameters for the GOALS sample measured from Pa$\alpha$, 8.4 Ghz, and CO (3-2). The authors also compile CO (3-2) and 1.4 GHz sizes for galaxies in the GOODS-N field. The diversity of measurements underscores the need for a homogeneous ALMA comparison. The bottom panel of Figure \ref{lir_morph} compares ISM diameter with $L_{\rm IR}^{\rm \scriptscriptstyle SF}/M_{\rm dust}$. The GOALS DSFGs are all much smaller than the high $z$ DSFGs. However, this does not translate to different $L_{\rm IR}^{\rm \scriptscriptstyle SF}/M_{\rm dust}$ or $L_{160}^{\rm \scriptscriptstyle SF}/L_{70}^{\rm \scriptscriptstyle SF}$ ratios. This echoes the result in \citet{lee2016}, where the authors find that local dust obscured galaxies, likely very compact, with $L_{\rm IR} = 10^{11}-10^{12}$ have similar dust masses and temperatures as their less obscured, more extended counterparts. Similarly, using SEDs output from hydrodynamical simulations, \citet{martinez2016} recently found that mergers and compactness alone are not able to reproduce the SEDs of local LIRGs, but that increasing the gas fraction was required. % 204 of these galaxies have $\fAGN<0.5$, and 25\% are isolated, 20\% are a close pair, and 54\% are interacting (we combine early and advanced stages into one category). 

However, without a large sample of measurements of galaxy size in the dust continuum (i.e., with ALMA), it is very difficult to distinguish between size and $M_{\rm dust}$ as the driver of the colder dust temperatures, since these effects are linked. Figure \ref{lir_morph} clearly demonstrates that the high $z$ galaxies are larger. If a galaxy can be represented as a central heating source surrounded by a disk of dust, then naturally a more extended disk will produce lower $T_{\rm dust}$, as the outskirts of the disk will see a diminished radiation field \citep[e.g.][]{misselt2001}. But, this is not a realistic description of most galaxies. Instead, to first order, both the SFR and dust densities are correlated with gas density, and so a mixed geometry of SFR regions within the ISM is more likely. In this geometry, the extent of the system doesn't affect the global effective dust temperature \citep{misselt2001,saf2015}.
How size and increased dust/gas mass are linked is also difficult to unravel. Locally, compaction can lead to a more efficient transformation of atomic gas into molecular gas \citep{larson2016}, so a decrease in size could be actually the trigger of increased molecular gas masses in many of the GOALS galaxies \citep{diaz2010,diaz2011}. Alternately, smaller size translates to a higher SFR surface density, which would boost $L_{\rm IR}^{\rm \scriptscriptstyle SF}$ without requiring a boost in $M_{\rm gas}$, leading to higher SFE at smaller size \citep{hayward2011,hayward2012}. These effects are observed locally and in simulations, but our small sample of high $z$ galaxies do not appear to show any of the same trends between small size and higher temperatures or $L_{\rm IR}^{\rm \scriptscriptstyle SF}/M_{\rm dust}$. These galaxies also have optical radii measurements from the CANDELS collaboration, and we find no correlation between optical radius measured in the observed frame $H$-band and $L_{\rm IR}^{\rm \scriptscriptstyle SF}/M_{\rm dust}$. On the other hand, \citet{scoville2015} argue that high $z$ galaxies have more turbulent ISMs, leading to compression in the ISM and enhancing the star formation efficiency per unit mass. This efficient mode of star formation can occur throughout the galaxy, so no obvious correlation between galaxy size and dust temperature may be expected.

Local galaxies can be resolved, allowing their ISM geometry to be measured in more detail than for high redshift galaxies \citep[e.g.,][]{barcos2016}. Given the quality and abundance of observations, it is more straightforward to link mergers with compact starbursts with dust heating in the GOALS samples \citep{diaz2010}. However, without resolved observations of the ISM in high redshift galaxies, we see no obvious link between mergers, size, $L_{\rm IR}^{\rm \scriptscriptstyle SF}/M_{\rm dust}$, and $T_{\rm dust}$ in our sample. Therefore, we must conclude that 
a galaxy's global far-IR/submm emission, parameterized through $L_{\rm IR}$, $T_{\rm dust}$, $M_{\rm dust}$, or $f_{\rm gas}$ may not tell observers anything about the ISM geometry, extent, or merger stage of that galaxy.

\subsection{The Effect of Increased Gas Fraction}
\label{sec:gas}
There are two likely explanations for the observed increase in $M_{\rm dust}$ with redshift. The first is that DSFGs at $z\sim1-2$ are simply more massive overall, that is, they have a higher stellar mass. $M_\ast$ is expected to broadly scale with $M_{\rm dust}$, since a higher $M_\ast$ in a DSFG implies a higher metallicity, supplying more metals to form dust in the ISM.

\begin{figure}
\includegraphics[width=3.3in]{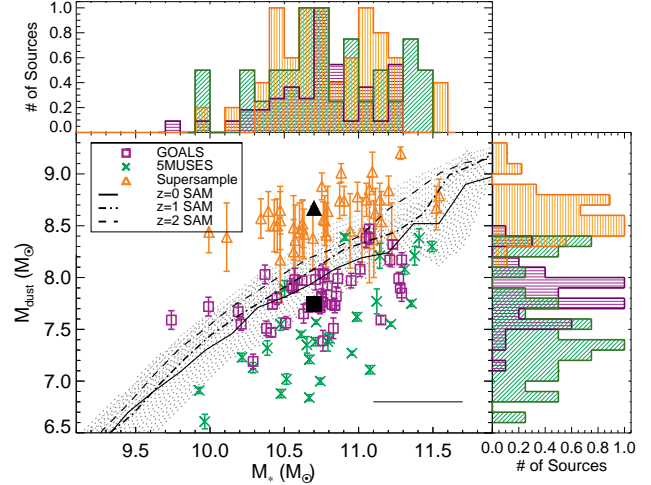}
\caption{We explore how much of the increase in dust mass is related to the evolution of the main sequence (MS) with redshift. The Supersample (orange triangles) span the same range of stellar mass as the GOALS galaxies (purple squares) and 5MUSES galaxies (green crosses). We overplot the predicted relationship between $M_{\rm dust}$ and $M_\ast$ for 3 different redshifts from the models of \citet{popping2016}. Although generally consistent with the lower redshift galaxies, the Supersample galaxies have significantly more dust than predicted by the models. We also predict what $M_{\rm dust}$ a galaxy with $\log M_\ast = 10.7 \,M_\odot$ would have at $z=0$ and $z=1.2$ based on the evolution of $f_{\rm gas}$ with $z$ and assuming a constant dust to gas ratio. The dark triangle ($z=1.2$) and square ($z=0$) demonstrate the predictions are consistent with the measurements. The bar in the lower right indicates the estimated errors on the 5MUSES and GOALS galaxies based on comparison with $K-$band derived stellar masses. If the 5MUSES galaxies are overestimated, this would explain the lack of consistency with the \citep{popping2016} models at $z\sim0$.\label{stellar_mass}}
\end{figure}

In Figure \ref{stellar_mass}, we compare $M_{\rm dust}$ with $M_\ast$. The three samples span the same range of $\log M_\ast\sim10-11.5\,M_\odot$. However, the Supersample have roughly an order of magnitude higher dust masses. The solid, dotted, and dashed lines indicate how the relationship $M_{\rm dust}$ v. $M_\ast$ is predicted to evolve with redshift using the semi-analytic models of \citet{popping2016}. We find an increase in the $M_{\rm dust}$ with redshift larger than predicted by those models. The Supersample is offset from the GOALS galaxies by $\sim0.7\,$dex, which is the same amount of offset seen in the bottom panel of Figure \ref{mdust_lir}. The \citet{popping2016} models shown in our comparison are based on a semi-analytic model of galaxy formation in a cosmological context, which includes a standard suite of physical processes (gas accretion and cooling, star formation, stellar feedback, chemical enrichment, etc). In addition, the \citet{popping2016} model includes self-consistent tracking of the main processes thought to produce and destroy dust in galaxies, including dust condensation in stellar ejecta, dust growth through accretion in the ISM, dust destruction by supernovae, and ejection of dust by stellar driven winds.  The much milder evolution of $M_{\rm dust}/M_\ast$ with redshift predicted by these models relative to our findings is interesting, as it indicates that one or more of the model ingredients needs to be revised.

%\citet{mckinnon2016b} simulate the dust content of galaxies with $M_\ast=10^6-10^{13}M_\odot$ from $z=0-2$ and find that dust mass depends on both $M_\ast$ {\it and} SFR:
%\begin{equation}
%\log \left(\frac{M_{\rm dust}}{M_{\rm dust,0}}\right)=\alpha \log\left(\frac{M_\ast}{10^{10}M_\odot}\right)+\beta\log\left(\frac{\rm SFR}{M_\odot{\rm yr}^{-1}}\right)
%\end{equation}
%where, for $z=0$, $\alpha=0.47$, $\beta=0.06$, and $M_{\rm dust,0}=4\times10^6M_\odot$. Even correcting for differences in cosmology, these parameters do not fit our low $z$ galaxies, but this is likely due to the fact that \citet{mckinnon2016b} are fitting to simulated galaxies down to $M_\ast = 10^6\,M_\odot$. We refit this relationship to our low $z$ and high $z$ samples separately. For the low $z$ sample, we find
%$\alpha=0.38,\beta=0.56,M_{\rm dust,0}=3.3\times10^6M_\odot$. Unlike in the larger simulated sample, our observed galaxies show a moderate dependence on {\it both} $M_\ast$ and SFR. For the high $z$ sample, we find $\alpha=-0.06,\beta=0.14,M_{\rm dust,0}=2.2\times10^8M_\odot$. This means that the high $z$ galaxies show very little correlation between $M_{\rm dust}$ and $M_\ast$, which is evident in Figure \ref{stellar_mass}. Again, the fitted parameters indicate an evolution in the relationship between these three properties with redshift.

The main sequence, which is the relationship between SFR and $M_\ast$, evolves with redshift \citep[e.g.][]{whitaker2012}. That is, for a given $M_\ast$, galaxies had a higher SFR at $z\sim1-2$ than today. A higher SFR 
can be tied to an increase in dust mass indirectly, as the majority of grain growth is predicted to occur in the ISM, and a more gas rich ISM will lead to higher dust masses and higher SFRs \citep[e.g.,][]{dwek1998,draine2003,santini2014,mckinnon2016a}. The increased dust mass in the Supersample might then be a natural consequence of the increase in gas fractions with lookback time. As mentioned above, submm data can also be converted to  $M_{\rm H_2}$, and we use this parameterization to calculate gas fractions $f_{\rm gas} = M_{\rm H_2} / (M_\ast + M_{\rm H_2}$). These gas fractions are consistent with what we derive using CO observations for several 5MUSES and Supersample galaxies \citep{yan2010,kirkpatrick2014}.

We also calculate the distance of these galaxies from the main sequence. The main sequence evolves with redshift, and it flattens at higher $M_\ast$. Therefore, we follow the method in \citet{scoville2017} to calculate the main sequence. We use the relationship between $M_\ast$ and SFR parameterized in \citet{lee2015} at $z=1.2$:
\begin{equation}
\log {\rm SFR_{MS}} = 1.72 - \log \left[1 + \left(\frac{M_\ast}{2\times10^{10} M_\odot}\right)^{-1.07}\right]
\end{equation}
and then we renormalize depending on redshift \citep{speagle2014}:
\begin{equation}
{\rm SFR_{MS}}(z) = \left(\frac{1+z}{1+1.2}\right)^{2.9}\times{\rm SFR_{MS}(z=1.2)}
\end{equation}
Then, we calculate ${\rm sSFR}={\rm SFR}/M_\ast$ for every source and for the main sequence (MS) at each redshift.

\begin{figure}
\includegraphics[width=3.3in]{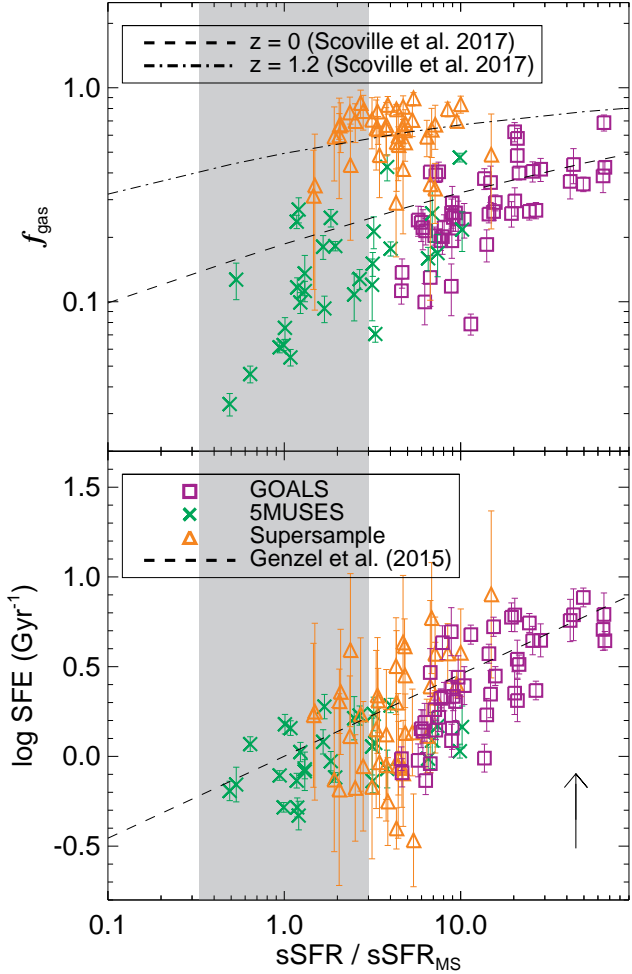}
\caption{{\it Top panel--} $f_{\rm gas} =  M_{\rm H_2} / (M_\ast + M_{\rm H_2}$) as a function of distance from the main sequence (grey shaded region). Starburst galaxies have higher gas fractions, and this increases with redshift. We overplot the predicted redshift evolution of $(1+z)^{1.84}$ from \citet{scoville2017}, and this evolution agrees with our samples. {\it Bottom panel--} Star formation efficiency as a function of distance from the main sequence. Now, there is no offset with redshift, in agreement with \citet{genzel2015}. In contract, the predicted evolution of $(1+z)^{1.05}$ measured by \citet{scoville2017} is shown as the arrow in the lower right corner for $z=0$ to $z=1.2$. \label{fgas}}
\end{figure}

We plot $f_{\rm gas}$ vs. $\Delta{\rm sSFR}={\rm sSFR}/{\rm sSFR}_{\rm MS}$ in the top panel of Figure \ref{fgas}. The Supersample is offset relative to the low $z$ samples, so that at a given $\Delta$sSFR, the Supersample sources have higher gas fractions, consistent with previous results in the literature \citep{magdis2012,genzel2015}. This means that massive galaxies at $z\sim1.2$ have $4-5\times$ higher gas fractions, and hence gas masses, since the stellar masses are roughly consistent. \citet{scoville2017} parameterize the evolution of the molecular gas mass as a function of redshift, $M_\ast$, and $\Delta$sSFR. In particular, the authors find that $M_{\rm H_2}$ evolves as $(1+z)^{1.84}$.
For $\log M_\ast = 10.7\,M_\odot$, and $z=0$, $z=1.2$, we calculate the gas masses based on Equation 6 in \citet{scoville2017} and overplot these predictions as the dashed and dot-dashed lines. The predicted redshift evolution of $(1+z)^{1.84}$ is consistent with our measurements, given the uncertainties in our stellar masses. 

This redshift evolution is best interpreted by also considering star formation efficiency, SFE = SFR/$M_{\rm H_2}$. In the case of a constant dust to gas ratio, SFE $\propto L_{\rm IR}^{\rm \scriptscriptstyle SF}/M_{\rm dust}$; massive dusty galaxies show little variation in metallicity, even at $z\sim1-2$, so a constant dust to gas ratio is a fair assumption \citep{magdis2012}. We plot SFE vs. $\Delta$sSFR in the bottom panel of Figure \ref{fgas}. The obvious correlation (Kendall's $\tau=0.45$) is similar to the well established dependence of the gas depletion timescale ($t_{\rm dep}={\rm SFE}^{-1}$) on sSFR \citep{saintonge2011,magdis2012,tacconi2013,sargent2014,huang2014}. We overplot the relation derived in \citet{genzel2015}, where the authors use CO observations to calculate molecular gas mass for 500 SFGs from $z=0-3$. There is no offset in any of our samples, consistent with the very mild redshift evolution between $t_{\rm dep}$ and $\Delta$sSFR found by \citet{genzel2015}. However, this is in contrast to \citet{scoville2017}, which measure an increase in ${\rm SFE}\propto(1+z)^{1.05}$ for a given stellar mass. We plot what this evolution looks like as the arrow in the bottom right corner for $z=0$, $z=1.2$ and $\log M_\ast = 10.7\,M_\odot$. 

The clear increase in gas fractions with redshift, without corresponding increases in SFE, suggests that the higher SFRs (corresponding to higher $L_{\rm IR}$) and higher $M_{\rm dust}$ of the Superample at $z\sim1-2$ are driven by an increase in the gas supply rather than an increase in SFE, $M_\ast$, or changing mode of star formation with cosmic time \citep{bouche2010,tacconi2013,dave2012,genzel2015,schinnerer2016}.
In fact, the bottom panel of Figure \ref{mdust_lir} demonstrates that at a given $L_{\rm IR}^{\rm \scriptscriptstyle SF}$ (SFR), SFEs are {\it lower} at $z\sim1-2$. %However, the ratio of gas mass to stellar mass ($f_{\rm gas}$ for a given $L_{\rm IR}$ does not. 
%normalization of this correlation evolves with redshift, similar to previous results in the literature \citep{magdis2012,genzel2015}. The Supersample galaxies for which we are able to measure gas
%fractions have an average redshift of 1.4. This means that massive galaxies at $z=1.4$, which are on the main sequence, have three times higher gas fractions than local MS DSFGs. 
This increase in gas supply is also predicted to by the underlying cause of the main sequence evolution, SFR/$M_\ast$, with redshift
\citep{bouche2010,tacconi2013,dave2012,genzel2015}. We can directly test whether an increase in $f_{\rm gas}$ fully explains the increase in $M_{\rm dust}$ using the relationship between $f_{\rm gas}$ and $z$, parameterized as $f_{\rm gas} = 0.1(1+z)^2$ \citep{geach2011}. If we assume a dust-to-gas ratio of 100 and $\log M_\ast=10.7\,M_\odot$, then at $z=0$, $\log M_{\rm dust}=7.75\,M_\odot$, and at $z=1.2$, $\log M_{\rm dust} = 8.67\,M_\odot$. These predictions are plotted as the dark filled symbols in Figure \ref{stellar_mass}, and they agree remarkably well with the GOALS and Supersample.

\section{Conclusions}
We have analyzed the far-IR/submm properties of three samples of infrared luminous galaxies. We combine GOALS and 5MUSES galaxies to form a low $z$ sample ($z<0.3$), and the Supersample comprises our high $z$ sample ( $z\sim0.5-2$). Mid-IR spectra are available for every source, and we utilize these spectra to identify sources with a hot dust continuum, likely due to an AGN, and remove them from the sample. We measure $L_{250}^{\rm \scriptscriptstyle SF}/L_{70}^{\rm \scriptscriptstyle SF}$, $L_{160}^{\rm \scriptscriptstyle SF}/L_{70}^{\rm \scriptscriptstyle SF}$, $M_{\rm dust}$, $f_{\rm gas}$ self-consistently for all galaxies. We find
\begin{enumerate}
\item $L_{250}^{\rm \scriptscriptstyle SF}/L_{70}^{\rm \scriptscriptstyle SF}$ is tightly correlated with $L_{160}^{\rm \scriptscriptstyle SF}/L_{70}^{\rm \scriptscriptstyle SF}$ for all galaxies and both ratios are sensitive to the strength of the ISRF, as parameterized by SED models. The low $z$ and high $z$ samples span the same range of colors, although for a given $L_{160}^{\rm \scriptscriptstyle SF}/L_{70}^{\rm \scriptscriptstyle SF}$, the high $z$ galaxies have a higher 
$L_{250}^{\rm \scriptscriptstyle SF}/L_{70}^{\rm \scriptscriptstyle SF}$, indicating that they have proportionally more cold dust. 
\item $L_{160}^{\rm \scriptscriptstyle SF}/L_{70}^{\rm \scriptscriptstyle SF}$ is correlated with $L_{\rm IR}^{\rm \scriptscriptstyle SF}$, but this relationship evolves with redshift so that at $z\sim1-2$, DSFGs are $\sim5\,$K colder.
\item There is a strong relationship between $L_{\rm IR}^{\rm \scriptscriptstyle SF}/M_{\rm dust}$ and $L_{\rm IR}^{\rm \scriptscriptstyle SF}$, and this relationship evolves with redshift. DSFGs at $z\sim1-2$ have $L_{\rm IR}^{\rm \scriptscriptstyle SF}/M_{\rm dust}$ ratios similar to low $z$ galaxies a factor of 5 less luminous. DSFGs also have higher dust masses than their local counterparts at the same luminosity.
\item The relationship between $M_{\rm dust}$ and $M_\ast$ also evolves strongly with redshift, so that DSFGs at $z\sim1-2$ have a $>$ factor of 5 higher $M_{\rm dust}$ than local DSFGs in the same $M_\ast$ range. %Because the low $z$ and high $z$ samples span the same $M_\ast$ range, a difference in $M_\ast$ cannot account for the higher $M_{\rm dust}$ at high $z$.
%While there is a broad correlation between $f_{\rm gas}$ and $L_{\rm IR}^{\rm \scriptscriptstyle SF}$ for all galaxies regardless of redshift, the high $z$ sample is still slightly offset from the low $z$ sample. %There is a correlation between $f_{\rm gas}$ (also calculated from submm fluxes) and $L_{\rm IR}$ %that broadly speaking, does not change with redshift. %That is, the intrinsic ratio of gas mass to stellar mass for a given $L_{\rm IR}$ does not evolve with time. 
This mirrors the measured increase in gas fraction with redshift from other studies.%As the low and high $z$ sources span similar $M_\ast$ ranges, the increase in $M_{\rm dust}$ with redshift cannot be explained by an increase in $M_\ast$ alone, but can also be attributed to an increase in the gas content of galaxies for a given stellar mass.
\item We observe a redshift-independent correlation between distance from the main sequence ($\Delta$sSFR) and star formation efficiency (SFE). The higher star formation rates (corresponding to higher $L_{\rm IR}$) in our $z\sim1-2$ sample are then fueled by an increase in their gas content, rather than an increase in SFE.
\item We see a tight redshift-independent correlation between $L_{\rm IR}^{\rm \scriptscriptstyle SF}/M_{\rm dust}$ and $L_{160}^{\rm \scriptscriptstyle SF}/L_{70}^{\rm \scriptscriptstyle SF}$, which is a proxy for $T_{\rm dust}$. We see no correlation between merger classification and either $L_{\rm IR}^{\rm \scriptscriptstyle SF}/M_{\rm dust}$ or $L_{160}^{\rm \scriptscriptstyle SF}/L_{70}^{\rm \scriptscriptstyle SF}$. This is also true of galaxy size. Then, the change to colder $T_{\rm dust}$ in high $z$ DSFGs can be explained simply by an increase in dust mass at fixed $L_{\rm IR}^{\rm \scriptscriptstyle SF}$, without requiring any change in the merger fraction or extent of the ISM, although this may be a related effect. 
\item In the subsample of galaxies for which we have size measurements, the large galaxy sizes at high $z$ and lack of a correlation between size and $L_{\rm 160}/L_{\rm 70}$ favor an ISM geometry in which the stars and dust are well mixed, so the expectation based on a central source surrounded by dust (that $T_{\rm dust}$ is inversely proportional to size) does not hold.
\item The far-IR/submm SED can be fully parameterized in terms of the observables $L_{\rm IR}^{\rm \scriptscriptstyle SF}$ and $L_{160}^{\rm \scriptscriptstyle SF}/L_{70}^{\rm \scriptscriptstyle SF}$ or, alternatively, the dust mass and the luminosity absorbed by dust (which is equal to $L_{\rm IR}^{\rm \scriptscriptstyle SF}$ unless dust self-absorption is non-negligible). As such, a galaxy's global far-IR/submm emission alone cannot be used to distinguish between ISM geometries. 
\end{enumerate}

We thank Caitlyn Casey and Gergo Popping for helpful conversations. We also thank the CANDELS collaboration for providing the visual merger classification catalogs for GOODS-N and GOODS-S. We thank the anonymous referee for helpful comments improving the scope of this paper.
AK, AP, and AS acknowledge NASA ADAP13-0054 and NSF AAG grants AST-
1312418 and AST-1313206. T.D-S. acknowledges support from ALMA-CONICYT project 31130005 and FONDECYT regular project 1151239.
The Flatiron Institute is supported by the Simons Foundation.

\clearpage

\appendix
\section{A. Rest Frame Photometry}
\label{appA}
We plot the IR/submm SEDs of the Supersample galaxies listed in Table \ref{properties}. We also plot the estimated rest frame 70, 160, 250\,$\mu$m photometry. For two galaxies, GN\_IRS49 and MIPS8543, we do not have enough coverage at $\lambda>200\,\mu$m to reliably estimate a rest frame $L_{250}$. The errors on the estimated rest frame photometry are largely attributable to the uncertainties on the templates \citep{kirkpatrick2015} used to derive the photometry.

\begin{figure*}[ht!]
\centering
\includegraphics[width=6in]{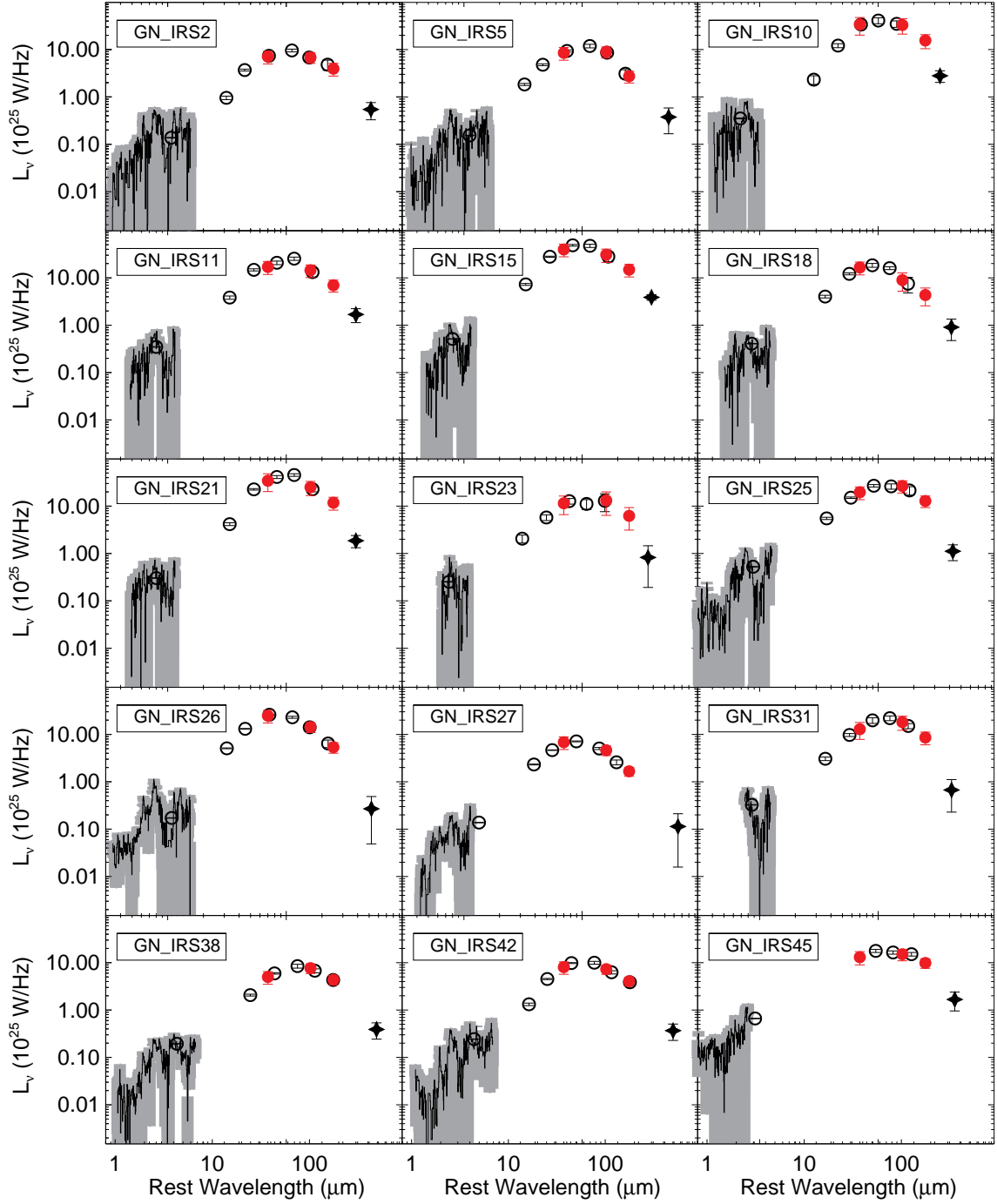}
\caption{We demonstrate how accurately the estimated rest frame 70, 160, 250\,$\mu$m photometry (red filled circles) matches the observed {\it Herschel} and {\it Spitzer} SEDs for the Supersample. We plot as the filled stars the ground based submm photometry used to calculate $M_{\rm dust}$.}
\end{figure*}

\setcounter{figure}{0}
\begin{figure*}
\centering
\includegraphics{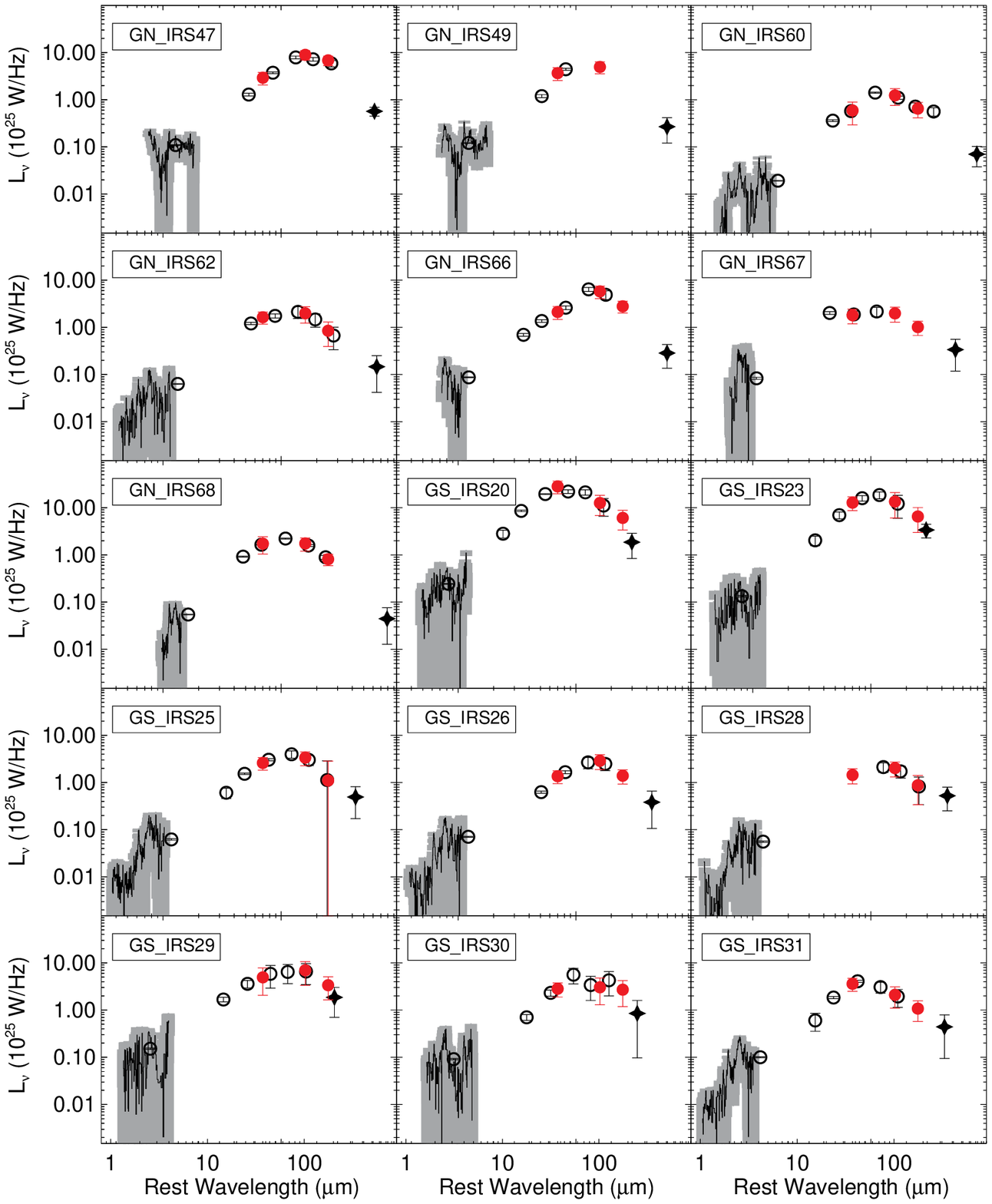}
\caption{\it Continued}
\end{figure*}

\setcounter{figure}{0}
\begin{figure*}
\centering
\includegraphics{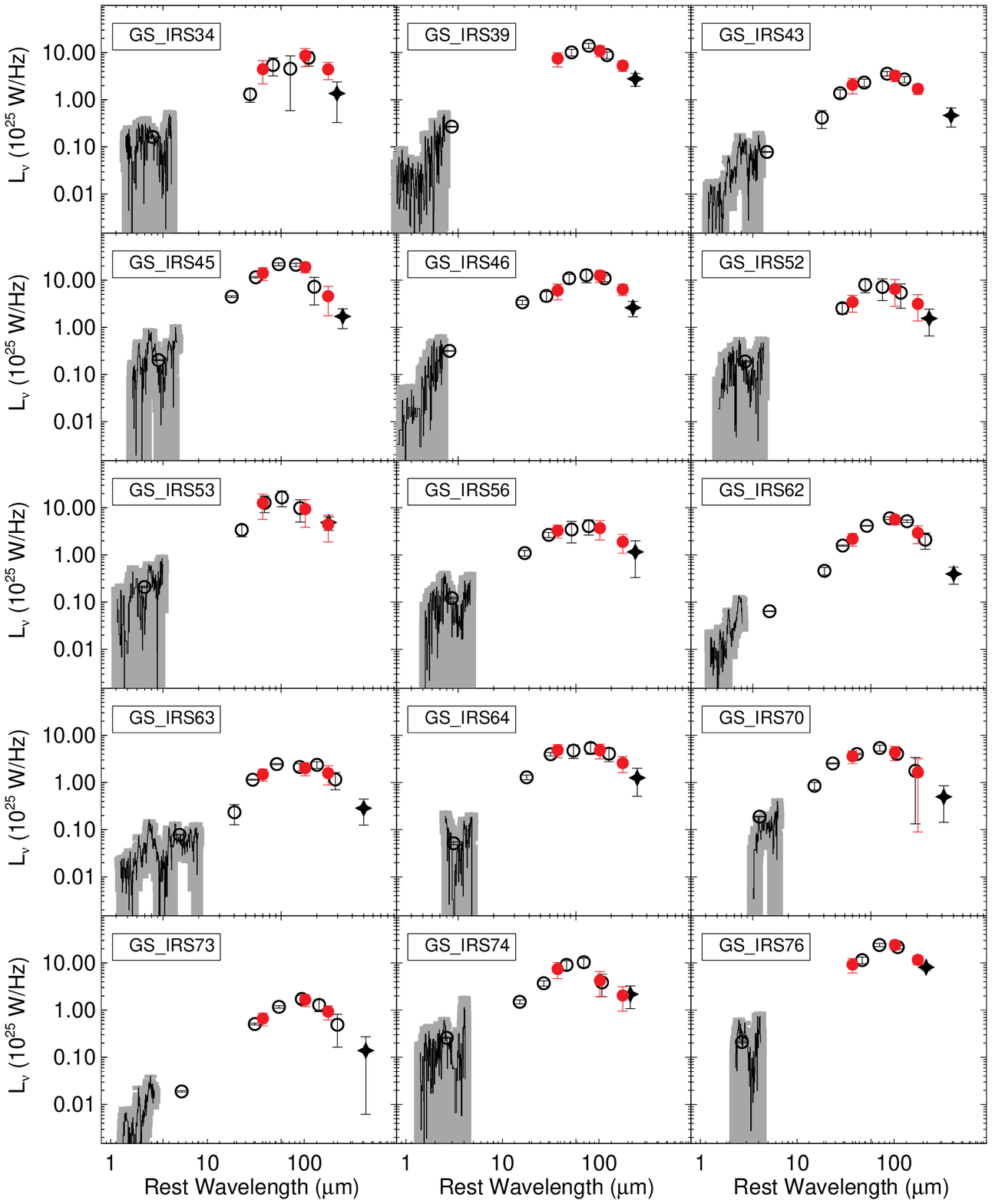}
\caption{\it Continued}
\end{figure*}

\setcounter{figure}{0}
\begin{figure*}
\centering
\includegraphics{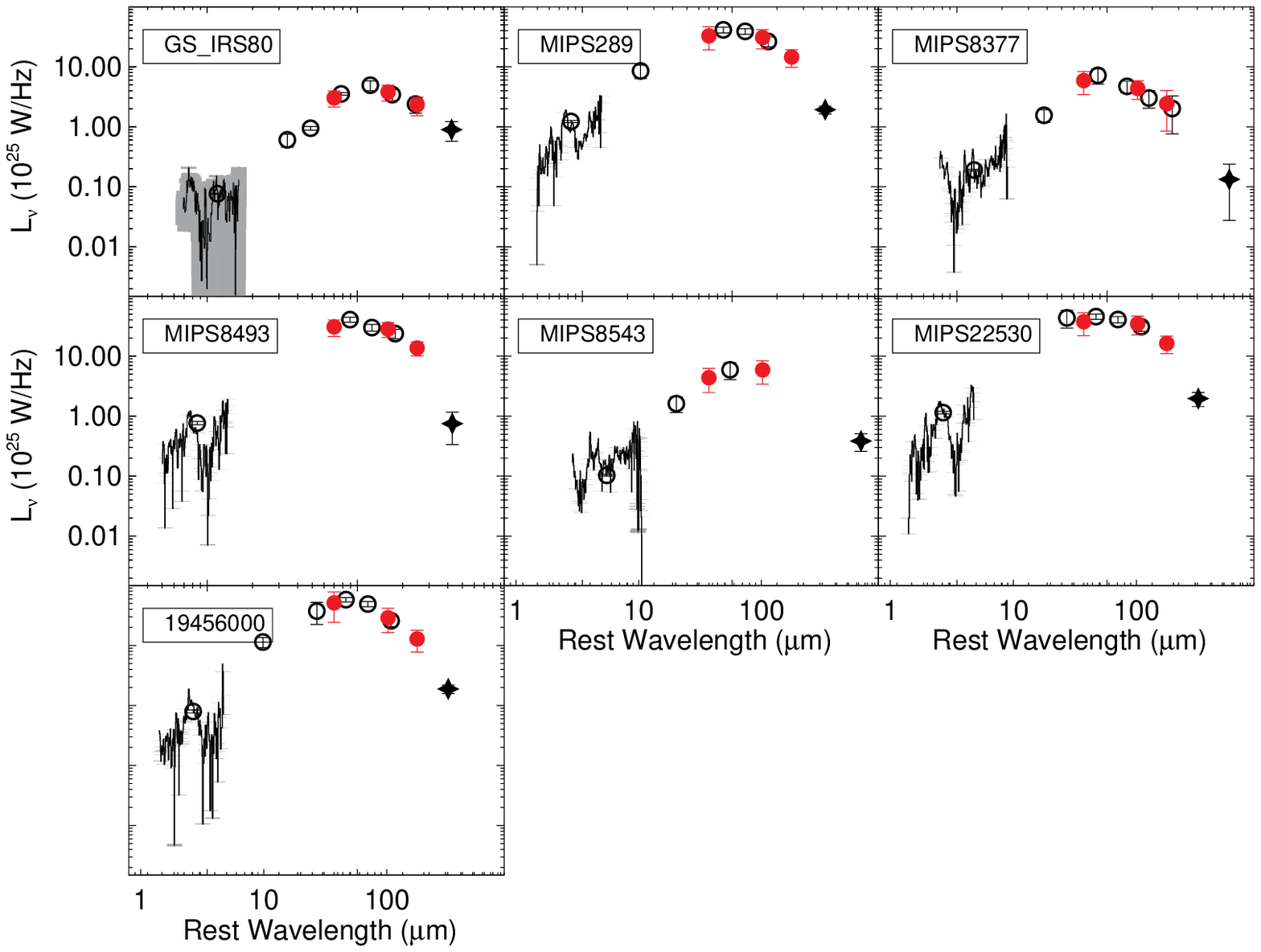}
\caption{\it Continued}
\end{figure*}

\clearpage
\section{B. Dust Mass Calculations}
\label{appB}
We explore how trends between $L_{\rm IR}^{\rm \scriptscriptstyle SF}$ and $M_{\rm dust}$ will change if we use different $T_{\rm dust}$ to calculate $M_{\rm dust}$.

Determining a dust temperature from the SED involves fitting a modified blackbody to the far-IR data. This temperature is then a luminosity-weighted temperature that may not represent the true temperature of the bulk of the dust mass. Instead, the effective dust temperature probes the mean ionization parameter $U$ ``seen" by the dust \citep{draine2007}. If a single modified blackbody is used, $T_{\rm dust}$ is measured using all far-IR photometry, but below $\lambda<100\,\mu$m, dust is being heated primarily in star forming regions, as opposed to the diffuse dust in the ISM heated by the interstellar radiation field \citep{persson1987,dunne2001}. Dust masses calculated from a one temperature modified blackbody can underestimate the true dust mass by at least a factor of 2 \citep{magdis2012}. The cold diffuse component is what makes up the bulk of the dust mass. $T_{\rm cold}$ is often measured by fitting a two temperature modified blackbody \citep[e.g.,][]{dunne2001,kirkpatrick2012,lee2016}, but this method requires full sampling of the far-IR SED or assumptions about $\beta$ and the warm dust component, not to mention that the model is unphysical. 

\begin{figure}
\plottwo{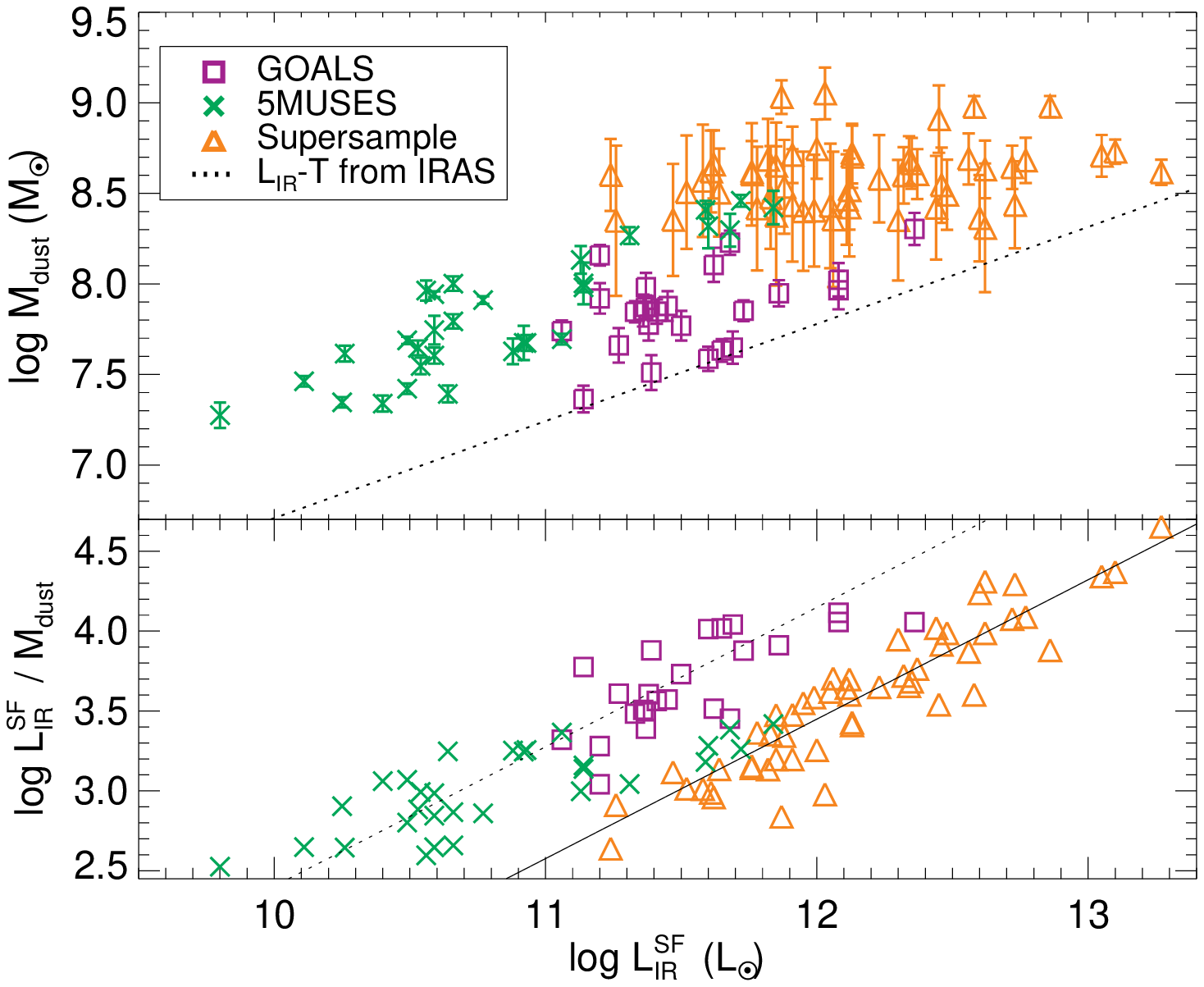}{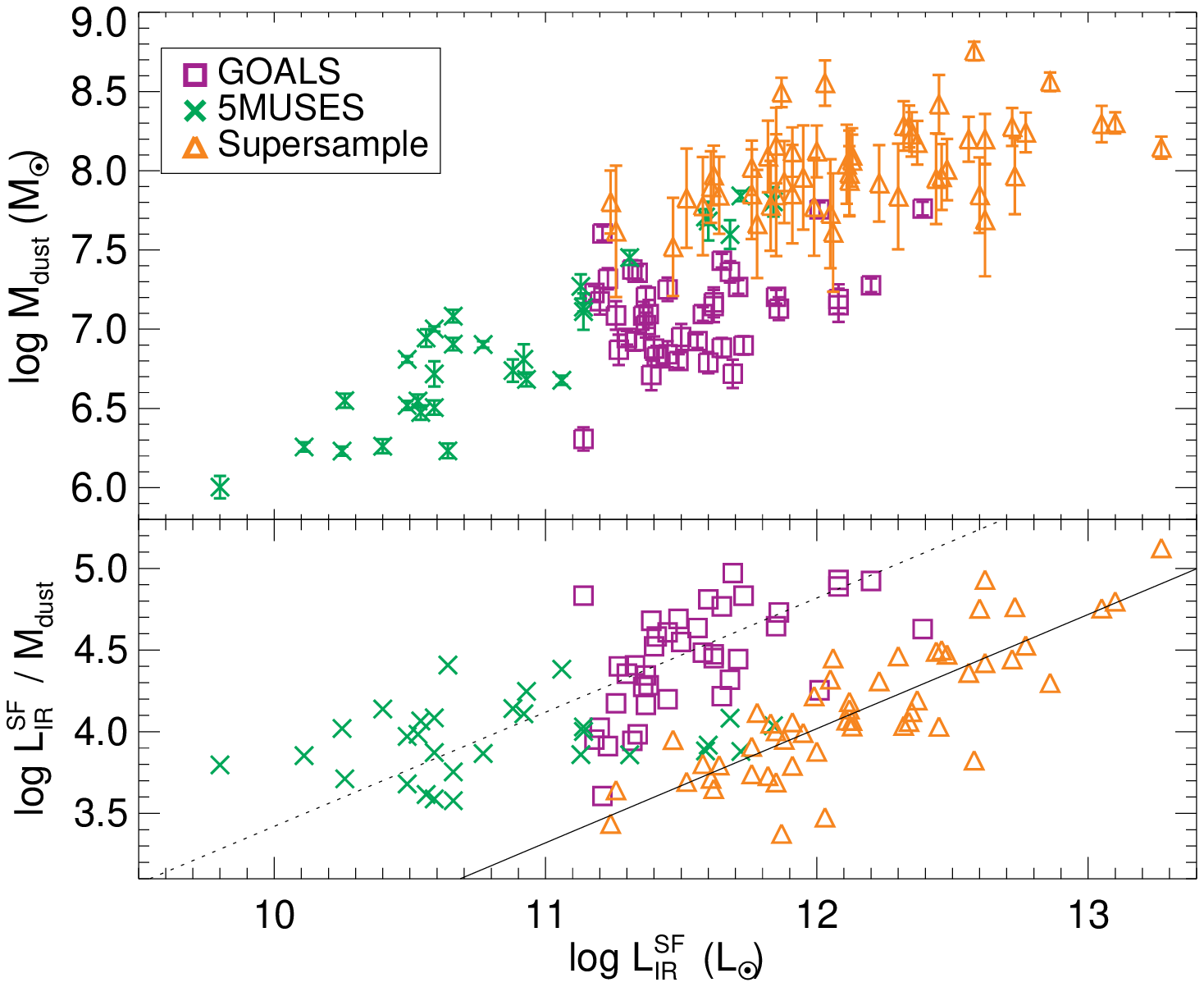}
\caption{$M_{\rm dust}$ v.\,$L_{\rm IR}^{\rm \scriptscriptstyle SF}$ with $T_{\rm dust}$ varying. {\it Left panel--} $T_{\rm dust}$ is approximated from a source's $L_{\rm IR}^{\rm \scriptscriptstyle SF}$. The dotted line shows how $M_{\rm dust}$ of the GOALS and 5MUSES galaxies will change when using the low $z$ $L_{\rm IR}-T_{\rm dust}$ relation from IRAS BGS sources \citep{chapman2003}. {\it Right panel--} $T_{\rm dust}$ and $\beta$ (used to derive $\kappa$) come from the best fitting IR template to each source from the library of \citet{dale2014}. In both panels, there is a clear evolution of $L_{\rm IR}^{\rm \scriptscriptstyle SF}/M_{\rm dust}$ with redshift, indicating that the method used to derive $M_{\rm dust}$ is not responsible for this trend. \label{Tdiff}}
\end{figure}

We now explore trends in dust mass when the dust temperature of the diffuse ISM is assumed to scale with increasing $L_{\rm IR}^{\rm \scriptscriptstyle SF}$, as this is consistent with what would be measured by fitting a one temperature modified blackbody to the SED \citep[e.g.,][]{blain2003,chapman2003,chapin2009,casey2012,lee2016}.
We calculate $T_{\rm dust}$ according to
\begin{equation}
\label{eq:LT}
T_{\rm dust}=0.47\times(L_{\rm IR}^{\rm \scriptscriptstyle SF})^{0.144}
\end{equation}
determined from 767 SPIRE 250\,$\mu$m selected sources spanning $z=0-5$ where $T_{\rm dust}$ was calculated from the peak wavelength of the SED \citep{casey2012}; we ignore the redshift evolution of the normalization for the time being. The left panel of Figure \ref{Tdiff} shows $M_{\rm dust}$ as a function of $L_{\rm IR}^{\rm \scriptscriptstyle SF}$. There is an overall trend towards increasing $M_{\rm dust}$ with increasing $L_{\rm IR}^{\rm \scriptscriptstyle SF}$, although for the the Supersample, $M_{\rm dust}$ is nearly constant over two orders of magnitude in $L_{\rm IR}^{\rm \scriptscriptstyle SF}$. The Supersample is offset 
from the GOALS and 5MUSES sample, with higher $M_{\rm dust}$ at similar $L_{\rm IR}^{\rm \scriptscriptstyle SF}$, confirming the trend in Figure \ref{mdust_lir}.

The GOALS and 5MUSES galaxies lie decidedly below the Supersample, and this is not explained by evolution in the $L_{\rm IR}-T_{\rm dust}$ relation with redshift. We can test for the effect of redshift by instead using the $L_{\rm IR}-T_{\rm dust}$ relation derived from local IRAS BGS sources, which we approximate as $T_{\rm dust}\propto (L_{\rm IR}^{\rm \scriptscriptstyle SF})^{0.137}$ \citep{chapman2003,casey2012}. When we apply the IRAS BGS relation to the 5MUSES and GOALS galaxies, they lie around the dotted line in Figure \ref{Tdiff}. Rather than bringing the low $z$ and high $z$ samples into better agreement, using two $L-T$ relations will make the dust masses {\it more} discrepant. %Now, galaxies at $z\sim1-3$ having an order of magnitude higher dust mass, or more, than local galaxies of the same $L_{\rm IR}$.

Alternately, we can calculate $T_{\rm dust}$ and $\beta$ through template fitting. The \citet{dale2014} library, which we use to approximate the far-IR colors of our galaxies in Figure \ref{s250_70_160}, has an effective $T_{\rm dust}$ and $\beta$ associated with each template. We fit the \citet{dale2014} library to all sources using a $\chi^2$ minimization technique. We then use the best fit template $\beta$ to scale $\kappa$ as

\begin{equation}
\kappa_\lambda = \kappa_{850} \times\left(\frac{850}{\lambda}\right)^\beta
\end{equation}

where $\kappa_{850}=0.15\,{\rm m}^2/{\rm kg}$ \citep{weingartner2001} and $\lambda$ is the submm wavelength we calculate dust mass at for each galaxy. We use $\kappa_\lambda$ and $T_{\rm dust}$ associated with the best fit template to calculate $M_{\rm dust}$. The results are shown in the right panel of Figure \ref{Tdiff}. Again, the same increase in $M_{\rm dust}$ with redshift for a given $L_{\rm IR}^{\rm \scriptscriptstyle SF}$ is seen. The only difference is that now, at a given $L_{\rm IR}^{\rm \scriptscriptstyle SF}$, the low $z$ galaxies have $\log L_{\rm IR}^{\rm \scriptscriptstyle SF}/M_{\rm dust}$ that is 0.8 dex below the high $z$ galaxies, slightly larger than the 0.7 dex offset in Figure \ref{mdust_lir}. We conclude that the observed evolution of $L_{\rm IR}^{\rm \scriptscriptstyle SF}/M_{\rm dust}$ with redshift is real and not a product of how we are deriving dust masses.

\end{document}